\documentclass[a4paper,fleqn,usenatbib,useAMS]{mnras}

\usepackage{graphicx}	
\usepackage{amsmath}	
\usepackage{amssymb}	
\usepackage{multicol}        
\usepackage{bm}		
\usepackage{pdflscape}	

\usepackage[T1]{fontenc}
\usepackage{ae,aecompl}
\usepackage{mathptmx}
\usepackage{float}
\newcommand{\kms}{\,km\,s$^{-1}$} 

\title[]{SDSS-IV MaNGA: Bayesian analysis of the star formation history of low-mass galaxies in the local Universe}

\author[S. Zhou]{Shuang Zhou$^{1}$\thanks{Contact e-mail: \href{mailto:szhou@mail.tsinghua.edu.cn}{szhou@mail.tsinghua.edu.cn}},
H.J. Mo$^{2,1}$,
Cheng Li$^{1}$,
M\'{e}d\'{e}ric Boquien$^{3}$,
Graziano Rossi$^{4}$
\\
$^{1}$Department of Astronomy, Tsinghua University, Beijing 100084, China\\
$^{2}$Department of Astronomy, University of Massachusetts Amherst, MA 01003, USA\\
$^{3}$Centro de Astronom\'{i}a, Universidad de Antofagasta, Avenida Angamos 601, Antofagasta 1270300, Chile\\
$^{4}$Department of Astronomy and Space Science, Sejong University, 209, Neungdong-ro, Gwangjin-gu, Seoul, South Korea
}
\date{Last updated 2015 May 22; in original form 2013 September 5}

\pubyear{2019}

\begin{document}
\label{firstpage}
\pagerange{\pageref{firstpage}--\pageref{lastpage}}
\maketitle

\begin{abstract}
We measure the star formation histories (SFH) of a sample of low-mass galaxies 
with $M_\ast<10^9M_\odot$ from the SDSS-IV MaNGA survey. 
The large number of IFU spectra for each galaxy are either combined to 
reach a high signal to noise ratio or used to investigate spatial variations. 
We use Bayesian inferences based on full spectrum fitting. Our analysis based on 
Bayesian evidence ratio indicates a strong preference
for a model that allows the presence of an old stellar population, 
and that an improper model for the SFH can significantly underestimate 
the old population in these galaxies. The addition of NIR photometry to the 
constraining data can further distinguish between different SFH model families 
and significantly tighten the constraints on the mass fraction in 
the old population. On average more than half of the 
stellar mass in present-day low-mass galaxies formed at least 8 Gyrs ago, while about 30\% within the past 4 Gyrs. Satellite galaxies on average 
have formed their stellar mass earlier than central galaxies. The 
radial dependence of the SFH is quite weak.     
Our results suggest that most of the low-mass galaxies have an early 
episode of active star formation that produces a large fraction of their 
present stellar mass.    
\end{abstract}



\section{Introduction}

 Star formation in galaxies is regulated by a wealth of complex 
physical mechanisms, such as the formation, growth and merger of dark 
matter halos, and the cooling and heating of baryon gas by radiative 
and feedback processes. Characterising the star formation history 
of observed galaxies represents, therefore, an important step in 
understanding how galaxies form and evolve. Despite being the most 
abundant type of galaxy in the universe, dwarf (low-mass) galaxies
remain elusive as far as their formation and evolution is concerned.
Their observed blue colour is usually taken as an indication that these 
galaxies are dominated by young stellar populations
\citep[e.g.][]{Kauffmannb2003}. By studying stacked 
spectra from the Sloan Digital Sky Survey (SDSS), \cite{Heavens2004} found a very 
different formation history between low- and high-mass galaxies. 
While galaxies with stellar masses smaller than $10^{10}$ M$_{\odot}$ 
could be well represented by a flat star formation rate over the past 
3 Gyr and a declining rate towards earlier epochs, 
more massive galaxies generally form most of their stellar masses 
earlier. Recent investigations about low-mass galaxies, however, 
challenge the stereotype that low-mass galaxies are all young.
For local group dwarf galaxies in which individual stars can be 
resolved, the star formation history can be obtained 
through `archaeological' age reconstruction 
\citep[e.g.][]{Mateo1998,Dolphin2005,Aloisi2007,
Tolstoy2009,Weisz2011,Annibali2013,Weisz2014a,Sacchi2016,Albers2019}.
The analysis based on the colour-magnitude diagram of resolved stellar 
populations in general suggests that most stars in dwarf galaxies 
were formed more than 5 Gyr ago. For example, \cite{Weisz2011} 
analysed the star formation histories (SFHs) of 60 nearby 
(D<4 Mpc) dwarf galaxies and found that these galaxies 
on average have formed half of their stars before  $z\sim 2$, 
regardless of their morphological types.
The existence of such an old stellar population is supported by 
other types of observations. Using the CANDELS survey, 
\cite{vanderWell2011} found a population of extreme emission 
line galaxies at redshift $z\sim1.7$, with a number density 
so high that they can contribute a significant fraction of the 
total stellar mass contained in present-day dwarf galaxies 
with masses between $10^8$ and $10^9$ M$_{\odot}$. 
These authors suggested that most of the stellar mass of 
these dwarf galaxies should have formed before $z\sim 1$.
\cite{Kauffmann2014} used the 4000 \AA\ break and H$\delta_A$ 
indices in combination with SFR/M$_*$ derived from emission line 
measurements to constrain the SFHs of a sample of SDSS galaxies with 
stellar masses in the range $10^8-10^{10}{\rm M}_{\odot}$, 
and concluded that galaxies with stellar masses smaller than 
$10^9{\rm M}_{\odot}$ are not all young but with half-mass 
formation times ranging from 1 to 10 Gyr. Similar conclusions 
have also been reached by spectral energy 
distribution (SED) modelling of a sample of blue 
compact dwarf galaxies with masses between $10^7$ and 
$10^9 {\rm M}_{\odot}$ \citep{Janowiecki2017} and a sample of HII galaxies 
\citep{Telles2018}. Stars in most of these galaxies are best described 
by two or more stellar populations, with the oldest population 
often dominating the stellar mass.

In agreement with with those observations, the empirical model presented 
in \citet{Lu2014} and \citet{Lu2015} predicts that dwarf galaxies 
in small dark matter halos ($M_h<10^{11}h^{-1}{\rm M_{\odot}}$) 
had a strong episode of star formation at $z>2$, producing 
significant amounts of old stars in them. However, such an 
old population is not predicted in many other models \citep[][]{Lim2017}.
SFHs of low-mass galaxies have also been investigated using hydrodynamical 
simulations. For example, \cite{Digby2019} analyzed the SFHs of a number of 
field and satellite dwarf galaxies in the APOSTLE and Auriga 
simulations, and found that the predicted star mass fractions of stars of 
different ages are quite different from those observed in real surveys.
\cite{Kimmel2019} analyzed about 500 dwarf galaxies in the FIRE-2 
zoom-in simulations and found that the cumulative SFHs of the simulated 
galaxies do not match those observed.

Clearly, accurate measurements of SFHs of low-mass galaxies can 
provide important constraints on galaxy formation models. 
However, the investigations so far have their limitations. 
For example, the number of dwarf galaxies for which stars can be 
resolved is very limited \citep{Weisz2011}, and so it is difficult to 
draw reliable statistical conclusions. Methods based on 
SED fitting \citep[e.g.][]{Janowiecki2017, Telles2018} 
can make use of galaxy photometry over a wide wavelength 
coverage, but they may lose spectral features that contain 
information about the SFH. This shortcoming may be remedied
by methods using galaxy spectra, but high SNR spectra are 
needed to probe the faint old stellar population. 
The stacking of spectra of individual galaxies has been   
used to achieve a high enough SNR for such analysis  
\citep[e.g.][]{Kauffmann2014}, but such stacking mixes 
the signals of individual galaxies.

Here we intend to make our contribution
by analyzing a sample of low-mass galaxies selected from the Mapping
Nearby Galaxies at Apache Point Observatory (MaNGA; \citealt{Bundy2015}).
With its integral-field spectroscopy (IFS), MaNGA provides a large 
number of integral-field unit (IFU) spectra of individual galaxies. 
These not only allow us to obtain high signal-to-noise composite spectra 
for individual galaxies, which is essential for constraining 
the SFH in detail, but also to study spatial variations 
of the SFH within individual galaxies. The large sample of low-mass 
galaxies also makes it possible to study how the SFH depends on 
other galaxy properties. In addition, MaNGA is designed to 
overlap as much as possible with other observations, which allows 
us to make use of information from other observations, thereby  
to get more observational constraints. Our analysis is based on 
our newly developed stellar population synthesis (SPS) code, 
Bayesian Inference for Galaxy Spectra (BIGS), which has been successfully 
used to constrain the IMF of MaNGA early-type galaxies \citep{Zhou2019}. 
BIGS fits the full composite spectrum of a galaxy and constrains 
its SFH along with other properties of its stellar population. 
The Bayesian approach provides a statistically rigorous way to explore 
potential degeneracies in model parameters, and to distinguish between different 
models through Bayesian evidence. Moreover, the flexibility of BIGS also 
allows us to add new observational constraints in our inferences.  

The paper is organised as follows.  In \S\ref{sec:data} we present our
data reduction process, including sample selection and spectral
stacking procedure. We then introduce the SPS model and the Bayesian 
approach used to fit galaxy spectra in \S\ref{sec:analysis}. Our 
main results are presented in \S\ref{sec:results}, and we discuss 
some potential uncertainties in \S\ref{sec:Uncertainties}. 
Finally, we summarize and discuss our results in \S\ref{sec:summary}. 
Throughout this work we use a standard $\Lambda$CDM cosmology 
with $\Omega_{\Lambda}=0.7$, $\Omega_{\rm M}=0.3$ 
and $H_0$ = 70 \kms Mpc$^{-1}$.

\section{Data} 
\label{sec:data}

\subsection{The MaNGA survey}

As one of the three core programmes in the fourth-generation Sloan Digital Sky Survey 
(SDSS-IV, \citealt{Blanton2017}), MaNGA aims to collect high resolution,  
spatially resolved spectra for about 10,000 nearby galaxies in the redshift 
range $0.01 < z < 0.15$ \citep{Yana2016,Wake2017}. MaNGA targets are selected 
from the NASA Sloan Atlas catalogue
\footnote{\label{foot:nsa}\url{ http://www.nsatlas.org/}} (NSA, \citealt{Blanton2005}),
and are chosen to cover the stellar mass range 
$5\times10^8 {\rm M}_{\odot}h^{-2} \leq M_*\leq 3 \times 10^{11} 
{\rm M}_{\odot}h^{-2}$. For each target, the MaNGA IFU covers a radius up 
to either $1.5 R_e$ or  $2.5 R_e$ ($R_e$ being the effective radius), to 
construct the “Primary” and “Secondary” samples, respectively \citep{Law2015}. 
MaNGA observes the selected galaxies with the two dual-channel BOSS spectrographs 
\citep{smee2013} on the Sloan 2.5~m telescope 
\citep{Gunn2006}, which provides simultaneous wavelength coverage
over $3600-10,300$ {\AA}, with a spectral resolution $R\sim2000$ 
\citep{Drory2015}. The spectrophotometry calibration of MaNGA is 
described in detail in  \cite{Yanb2016}, while the initial performance 
is given in \cite{Yana2016}. Raw data from MaNGA are calibrated and reduced 
by the Data Reduction Pipeline (DRP; \citealt{Law2016}) to produce spectra with 
a relative flux calibration better than $5\%$ over more than $80\%$ of the 
wavelength range \citep{Yana2016}. In addition, MaNGA provides measurements 
of stellar kinematics (velocity and velocity dispersion), emission-line 
properties (kinematics, fluxes, and equivalent widths), and spectral 
indices for each spaxel through the MaNGA Data Analysis Pipeline \citep[DAP;][]{Westfall2019,Belfiore2019}.

\subsection{UKIDSS}

Near-infrared (NIR) photometric data are commonly used to trace the 
stellar mass of galaxies, which can be compared with the mass 
estimated from the stellar population synthesis modeling of the optical 
spectra provided by MaNGA. As described in \cite{Yanb2016}, the MaNGA 
targets are chosen to overlap as much as possible with the United Kingdom 
Infrared Telescope (UKIRT) Infrared Deep Sky Survey (UKIDSS)  
footprint. UKIDSS uses the Wide Field Camera (WFCAM) on the 3.8 m United
Kingdom Infra-red Telescope (UKIRT,\citealt{Casali2007}), providing ZYJHK images over a 
large sky coverage. The basic information of the survey can be found 
in \cite{Lawrence2007}, the photometric system is described in \cite{Hewett2006}, and the calibration is described in \cite{Hodgkin2009}. 
The UKIDSS data is reduced by the official pipeline and the science products 
are released through 
the WFCAM Science Archive\footnote{\label{foot:WSA}\url{http://wsa.roe.ac.uk/}}
(hereafter WSA, \citealt{Hambly2008}). 

\subsection{Sample selection}

The galaxy sample used here is selected from the internal data release
of MaNGA, the MaNGA Product Launch 7 (MPL-7), 
which includes a total of 4,621 unique galaxies and has been 
made public available together with the SDSS fifteen data release 
\citep[SDSS DR15][]{Aguado2019}. 
We select a set of the least massive galaxies ($M_*<10^9$ M$_{\odot}$) 
according to their total stellar masses given by NSA. 
During the selection, we exclude galaxies with apparent problems 
in MaNGA DRP and DAP data processing. Galaxies that host AGN or with 
severe sky line contamination at the red end of the spectra 
are also excluded. After this selection process, we obtain a sample 
of 254 low-mass galaxies.

We then cross-match galaxies in this sample with data from WSA, 
selecting galaxies that have measurements of SDSS u, g, r, i, z and 
UKIDSS Y, J, H, K band magnitudes. This cross-match yields a total of 
752 galaxies in MaNGA MPL7 and 22 of them are in the least massive 
sample. Considering potential differences between SDSS and MaNGA,  
such as flux calibrations,  we convolve the spectrum of a galaxy 
obtained by stacking its spaxels within 1~$R_e$ (see below) 
with the SDSS filters to derived its MaNGA $(g-r)$ colour. 
We only select galaxies whose MaNGA $(g-r)$ colours are 
within 0.05 mag of the $(g-r)$ colours listed in WSA. 
This yields a final sample of 19 low-mass galaxies with both 
optical and NIR photometry. 

In what follows, we use 
the sample of 254 low-mass galaxies selected from MaNGA
to investigate the statistical properties of the SFHs of these 
galaxies, and use the 19 low-mass galaxies with photometry from UKIDSS 
to study additional constraints on the SFHs from the NIR photometry.

\subsection{Spectral stacking}

The original spectra provided by MaNGA DRP have typical $r$-band 
SNRs of $4-8$ {\AA}$^{-1}$ toward the outer radii of galaxies \citep{Law2016}. 
For the low-mass galaxies considered here, the SNR can be as low as 
2 {\AA}$^{-1}$ due to their relatively low surface brightness. 
One thus needs to combine spectra in each IFU plate
of each individual galaxy to obtain a stacked spectrum with 
a sufficiently high SNR.  

We use two kinds of stacked spectra of every individual galaxy. 
First, to study the global SFHs of individual galaxies and the overall variations of 
SFH from galaxy to galaxy, we bin spaxels with elliptical annuli inside one $R_e$ of each galaxy to form 
a single spectrum. Second,  to study the variations of the 
SFH within a galaxy, we divide spaxels of the galaxy into three 
radial bins, $(0.0-0.3)R_e$, $(0.3-0.7)R_e$ and $(0.7-1.2)R_e$, 
according to their normalised radii of elliptical annuli 
given in MaNGA DAP, and stack the spectra within 
individual radial bins. These radial bins are similar to those
used in related MaNGA studies, such as \cite{Zheng_etal2019}.

The stacking procedure used here is similar to that in \cite{Zhou2019}, 
and we refer the reader to that paper for details.
As discussed in the MaNGA DAP paper \citep[see][for details]{Westfall2019},  
the SNR of stack spectra deviates from the simple noise propagating formula, 
because the spaxels provided by DRP are not fully independent.
Here we use the 
correction term given by \cite{Westfall2019} to account 
for the covariance between spaxels and to estimate the SNR. 
The correction term can be written as 
\begin{equation}
n_{\rm real}/n_{\rm no covar}= 1 + 1.62 \log(N_{\rm bin})
\end{equation}
where $N_{\rm bin}$ is the number of spectra used in the stacking, 
$n_{\rm real}$ and $n_{\rm no covar}$ are the corrected noise 
vectors and noise vectors that assuming no covariance between 
pixels (namely, those generated from the simple noise propagating 
formula) respectively. With this correction, the typical SNR of 
the stacked spectra is around 40 pixel$^{-1}$.

\section{Analysis}
\label{sec:analysis}

The inferences of stellar population properties from galaxy spectra 
can be achieved by comparing stellar population models with the 
observed spectra. In practice, two complementary approaches, 
absorption-feature modeling and full-spectrum fitting, have been 
widely used. The full spectrum in principle contains more information, but the information can be fully used only when 
both the continuum shape and spectral features are all modelled
accurately, and when the SNR of the spectra is sufficiently high.
In contrast, absorption features can be selected to have the greatest 
sensitivity to the main parameters of interest, such as age, 
metallicity and $\alpha$ enhancement \citep[][]{Worthey1994}. 
Apparently because of the short wavelength window, absorption features 
may avoid influences from spectral regions that are not properly 
described by the model. However, the 
shortcoming is that some important information contained in 
the rest of the spectrum may be missed. In this paper, we adopt the 
full spectrum fitting method using Bayesian statistics to infer the SFH 
of individual galaxies and its co-variance with other properties.

\subsection{The spectral synthesis model}

To accurately model galaxy spectra, proper templates that meet the 
resolution and wavelength coverage of the data are crucial.
Several popular codes are available to model simple stellar 
populations (SSPs) of given ages and metallicity, including  
BC03 \citep{BC03}, M05 \citep{Maraston2005}, and E-MILES \citep{Vazdekis2016}.
These SSP models can be combined with an assumption of the 
SFH of a galaxy to predict its spectrum.  
Different SSP models are based on different stellar templates and 
isochrones, and thus have their own merits and shortcomings. 
Given that the wavelength range of the original MaNGA data is 
$3600-10300$ {\AA}, and the median redshift of MaNGA galaxies 
is $z\sim$0.03, we select models that have uniform wavelength 
coverage to $\sim 9000$ {\AA}. In addition, the UKIDSS data 
require an extension of the coverage to $\sim 2.5\,{\rm \mu m}$. 
With these considerations, we decide to adopt the 
E-MILES\footnote{\label{foot:emiles}\url{http://miles.iac.es/}}
model.

The MILES models first presented in \cite{Vazdekis2010} are constructed with the MILES \citep{miles2006MNRAS} stellar library. Later on, EMILES models extend MILES both bluewards and redwards  using CaT \citep{Cenarro2001},
Indo-U.S. \citep{Valdes2004} and the IRTF stellar library \citep{Cushing2005,Rayner2009}. With all these stellar templates combined, the E-MILES SSP 
spectra cover the wavelength range from 1680.2 {\AA} to 
$5{\rm \mu m}$ with a moderately high spectral resolution. In 
particular, the SSPs reach a resolution of 2.51 {\AA} (FWHM)
over the range from 3540 {\AA} to 8950 {\AA}, which covers the 
main portion of the spectral range of MaNGA. The spectral 
resolution decreases towards longer wavelengths, but is 
sufficient for photometry calculations.

The E-MILES model is computed for several IMFs. 
As our focus is on low-mass galaxies, which tend to 
have a bottom-light IMF \citep{LHY2018}, we choose the model 
constructed with the Chabrier IMF. Moreover, E-MILES provides two 
sets of isochrones, the Padova+00 isochrones \citep{Girardi2000} 
and the BaSTI isochrones 
\footnote{\label{foot:BaSTI}\url{http://www.oa-teramo.inaf.it/BASTI}}. 
The impact of using different isochrones has not been 
investigated extensively in the literature. For MaNGA galaxies,
\cite{Ge2019} found that using different isochrones 
does not lead to significant changes in the fitting quality of 
observed spectra. Using mock spectra, however,  
the authors found that the Padova+00 model works better at 
low  metallicity ([Z/H]<-1.0), while the BaSTI model works better  
for galaxies of higher metallicity. Since low-mass galaxies
in general are metal poor, we use E-MILES templates with 
Padova+00 isochrones.

\subsubsection{Star formation history}

\begin{figure*}
\centering
\includegraphics[width=1.0\textwidth]{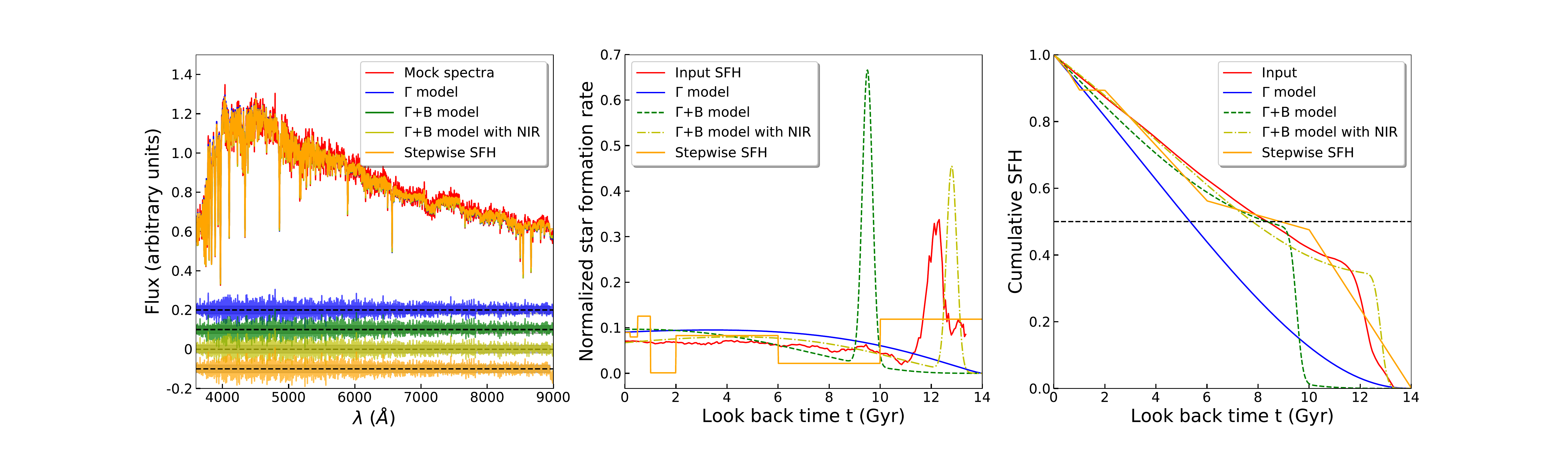}
\caption{An example of fitting mock spectra with different SFH models. 
The red line in the right panel shows the SFH used to generate the mock spectrum. 
White noise is added to the mock spectrum so that SNR=40. 
Blue solid and green dash lines are the best-fit results of the mock 
spectrum from 3800{\AA} to 8900 {\AA} using the $\Gamma$ and $\Gamma$ +B
SFH models, respectively, while the orange solid lines show the best-fit 
results from the stepwise model. Result 
of the $\Gamma$+B model using the optical spectrum plus the 
$(g-K)$ colour is shown as yellow dash-dotted lines.
Residuals of the best-fit spectra 
are shown at the bottom of the left panel, with zero points of the 
blue, green and orange lines shifted by 0.2,0.1,-0.1, respectively.
}

\label{fig:SFH_example}
\end{figure*}

\begin{figure*}
\centering
\includegraphics[width=1.0\textwidth]{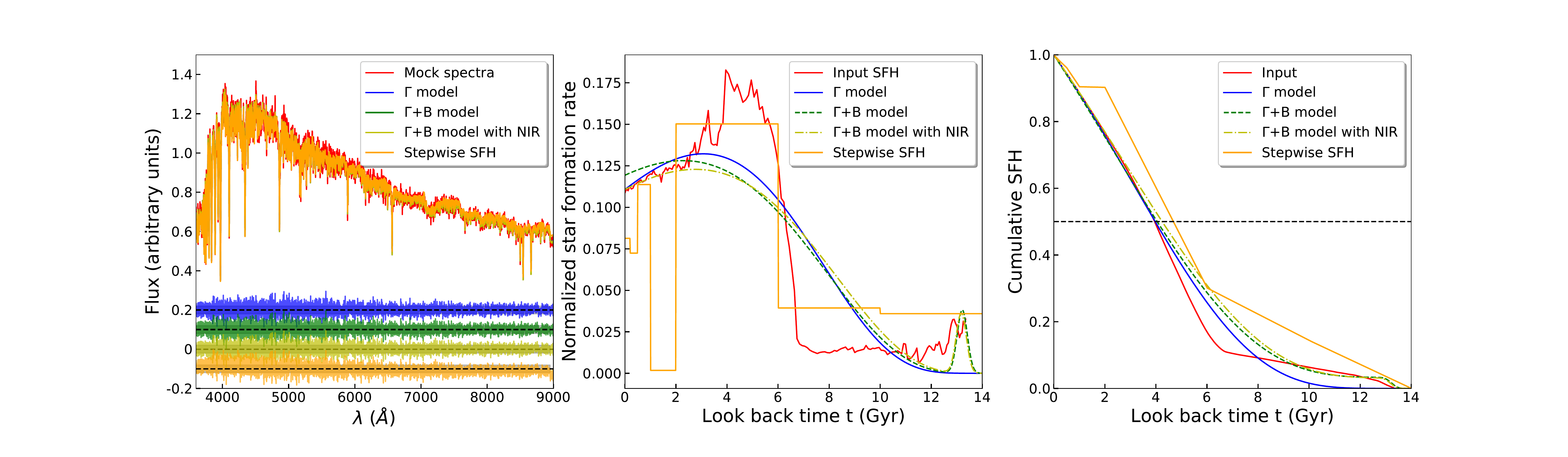}
\caption{An example similar to Fig.~\ref{fig:SFH_example}, but for a mock SFH that 
does not contain a significant early burst.}
\label{fig:SFH_example2}
\end{figure*}

\begin{figure*}
\centering
\includegraphics[width=1.0\textwidth]{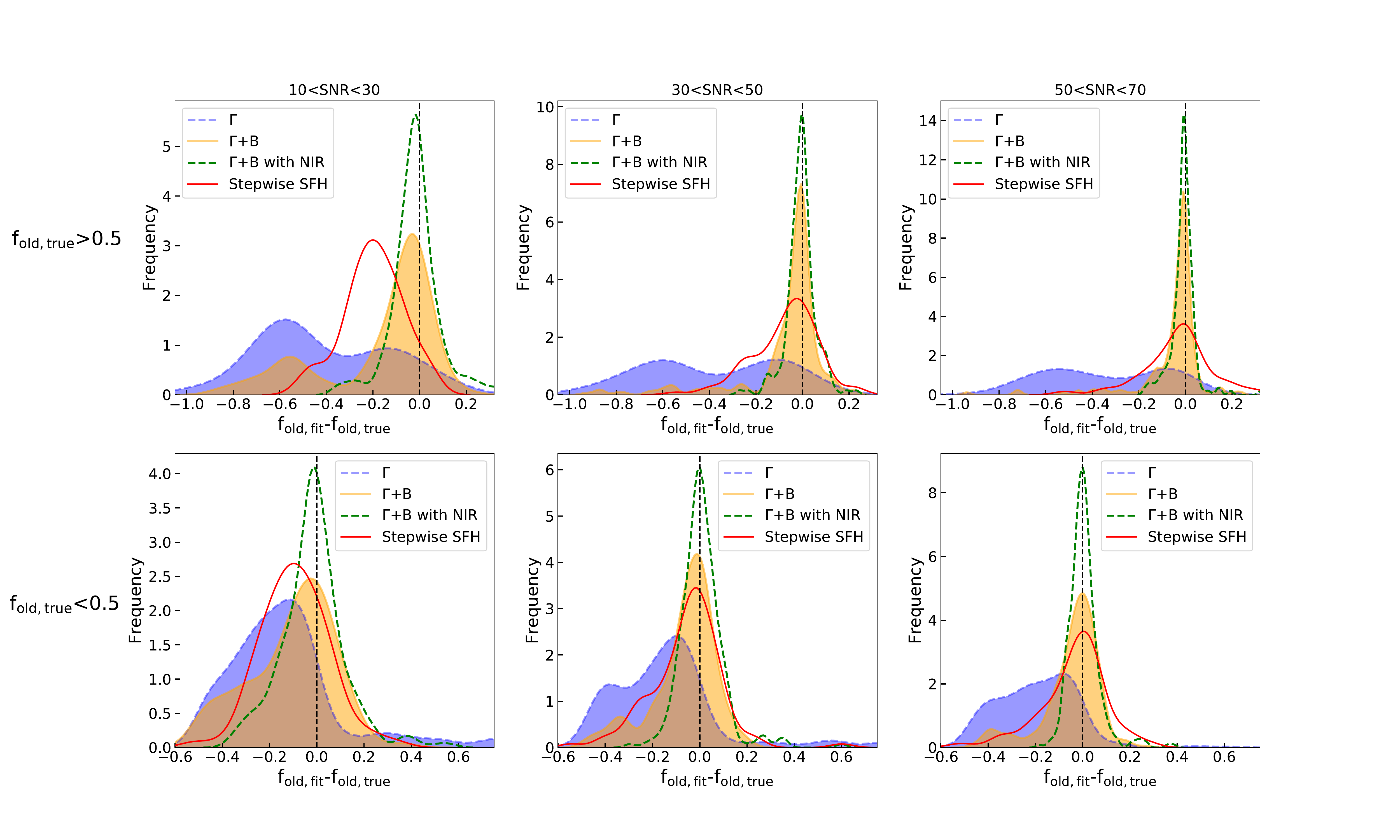}
\caption{The distribution of the difference in the mass 
fraction of the old stellar 
population (defined to be stars with ages larger than 8 Gyrs) 
between the best-fit model and the input SFH. The three columns 
are results for mock spectra with different levels of SNR, 
as indicated. Upper panels are results from mock spectra with 
high fraction (>0.5) of old stellar populations (>8 Gyr), 
similar to the one shown in Fig.~\ref{fig:SFH_example}, while 
the lower panels are for those with a lower fraction (<0.5) 
of old stars, similar to the one shown in Fig.~\ref{fig:SFH_example2}. 
The four distributions are results obtained from different SFH models, 
as indicated. Probability distributions are each normalized to 1.}

\label{fig:SFH_mock_sample}
\end{figure*}

In general, the details of the star formation history 
of a galaxy can be very complicated. However, limited by 
data quality, the observed galaxy spectra may be modeled  
in terms of a number of important stellar populations.
For low-mass galaxies, 
observations based on resolved stars \citep[e.g.][]{Weisz2011} 
have shown that the SFHs of different galaxies are quite similar 
over most of the cosmic time, while differences are only seen in 
the recent few Gyrs. These observations also indicate that
low-mass galaxies generally had enhanced star formation
in the early universe ($z>1$) where they formed more than 
half of their stars, and that some of them have gone through  
complex star formation histories during the recent 1 Gyr. 
Such two-phase star formation is also seen in empirical models 
such as that of \cite{Lu2015}.

Traditionally, SFHs of galaxies have been 
modelled with two distinct approaches: a parametric approach 
that assumes a functional form specified by a small number of parameters, 
and a non-parametric approach that models the SFH as a histogram
(a step-wise function) in a number of time bins. In 
our analysis, we use both approaches and make comparisons between them.

The parametric model adopted here is motivated by the 
empirical model of \cite{Lu2015}, who found that the SFH of a dwarf galaxy 
can be well represented by a burst in the early universe followed by a 
continuous SFH. To describe such a SFH, we model a smooth component 
using a $\Gamma$ function:
\begin{equation}
\Psi(t)=\frac{1}{\tau\gamma(\alpha,t_0/\tau)} 
\left({t_0-t\over \tau}\right)^{\alpha-1}
    e^{-(t_0-t)/\tau}\,,
\label{gamma-sfh}
\end{equation}
where $t_0-t$ is the look-back time,
and $\gamma(\alpha,t_0/\tau)\equiv \int_0^{t_0/\tau} x^{\alpha-1}e^{-x}\,dx$.
The flexibility of the function allows cases where a galaxy 
is dominated by old stars (with both $\alpha$ and $\tau$ small) 
or dominated by younger populations (with both $\alpha$ and 
$\tau$ large). On top of this continuous SFH, we include 
an additional burst component to mimic the old stellar 
component in  dwarf galaxies seen in resolved stars. 
The burst is specified by two free parameters:
$t_b$ describing when the burst occurs, and $f_{b}$ 
specifying the relative fraction of stellar mass formed in the burst.
Thus we consider the following two SFH model types:
\begin{itemize}
    \item $\Gamma$ model: SFH given by equation (\ref{gamma-sfh}); 
    \item $\Gamma$+B model: SFH given by equation (\ref{gamma-sfh}) plus a burst.
\end{itemize}

Non-parametric models are more flexible. In real applications, however, the 
flexibility depends strongly on the number of time bins used, 
and the accuracy of the inference that can be achieved is still limited by data quality. 
In addition, using too many time bins will lead to degeneracy in 
the solution and increase the computational time for sampling the posterior 
distribution. \cite{Panter2007} used 11 time bins in their 
analysis of SDSS galaxies, but found the model to be too ambitious 
for most galaxies. To model the color-magnitude diagram (CMD) based on 
resolved stars, \cite{Weisz2011} used a 6-step model to describe 
the SFH of local dwarf galaxies. In our analysis, we adopt
a stepwise SFH model similar to \cite{Weisz2011}.  
The model is described by the average star formation rates 
in 7 time intervals: $0\to 0.2$, $0.2\to 0.5$, $0.5\to 1.0$, $1\to 2$, 
$2\to 6$, $6\to 10$ and $10 \to 14$ Gyr. 
Since only the relative fraction of stars formed 
in each interval change the spectral shape, the 
model is specified by six free parameters.

We test the validity of these three forms of SFH using mock 
spectra generated 
from theoretical SFHs from the empirical model of \cite{Lu2015}. 
To do this, we first convolve the theoretical 
SFHs with the E-MILES SSPs to obtain the corresponding noise-free 
composite spectra. Different levels of Gaussian noise are then added to 
the spectra to mimic real observations.
An example of fitting such a mock spectrum with SNR=40 is shown in 
Fig.~\ref{fig:SFH_example}. 
Note that, in Fig.~\ref{fig:SFH_example}
and other figures, the burst component 
of the $\Gamma$+B model is represented by a Gaussian 
peak with finite width, even though it is modelled as a single SSP
in the fitting. As one can see, although both the 
$\Gamma$ and $\Gamma$+B models can give a reasonable fit to 
the mock spectrum, the former fails to recover the input SFH.
On the other hand, the $\Gamma$+B model has the flexibility to 
recover the early burst component, although the exact burst time is not reproduced. 
Similar to $\Gamma$+B, the step-wise SFH can also reproduce the 
early burst component, although the specific shape of the SFH is not 
accurately reproduced due to the time resolution used.

For comparison, we also show the result 
obtained using both the optical spectrum and $(g-K)$ colour to 
constrain the model (see \S\ref{ssec_fittingproc}). 
In this case the recovered SFH (the yellow curve) closely 
matches the input SFH, with approximately the same burst 
strength and time. Fig.~\ref{fig:SFH_example2} shows another example in which 
the mock SFH does not contain any significant burst.   
Here the inclusion of a burst component in the model and the NIR 
photometry in the constraint does not lead to any significant changes 
in the inferred SFH, as the $\Gamma$ function already has the 
flexibility to approximately describe this kind of SFHs. 
From the cumulative plots shown in the right panels of 
Figs.~\ref{fig:SFH_example} and \ref{fig:SFH_example2}, 
we can see that all models, except the $\Gamma$ model, predict 
similar half-mass formation times and old stellar fractions,
although the predicted shapes of the SFH are quite different 
in early times. Because of this, we will use the old stellar fraction 
and half mass formation time to characterize the old population, 
without paying much attention to the exact shape of the derived SFH.

To compare the true to the estimated physical properties,
we analyze 2,000 such mock spectra randomly chosen from the  
empirical models of \cite{Lu2015} which SNR ranging from 10 to 70. 
The differences in the mass fraction of the old stellar population 
(with stellar age >8 Gyrs) between the best-fit and input SFHs 
are shown in Fig.~\ref{fig:SFH_mock_sample}. It is seen that, the 
$\Gamma$ model systematically underestimates the mass fraction of 
the old population, regardless of the SNR. 
The $\Gamma$+B model can well recover the input old fraction, with 
some exceptions of underestimation at low SNR. 
The stepwise model also recovers the old fraction well, 
but there is a systematical underestimate at low SNR. 
In addition, including the NIR photometry as an additional constraint 
can improve the accuracy of the derived mass fraction, especially 
when the SNR is low. Although these test results are for ideal cases, 
where the stellar populations are perfectly described by the the SSPs, they do indicate that
the $\Gamma$+B model and the step-wise model can both recover the 
the old stellar population predicted by the empirical model 
in some low-mass galaxies, although the details of the 
SFH may not be modeled accurately.

In what follows, we will apply these SFH models to MaNGA spectra 
and examine whether the data prefer a particular SFH model, 
and how model inferences are affected by the assumption of 
the SFH.

\subsubsection{Dust attenuation}

Dust attenuation can also affect the inferred SFH. Since dust absorption is 
more significant at shorter wavelengths, a stellar population that 
contains more dust can mimic an older and/or more metal-rich population, 
producing the well known age-metallicity-dust degeneracy. Thus,  
dust attenuation has to be properly taken into account in order to 
make unbiased inferences from the observed spectra.
In practice, dust attenuation is usually treated as an additional model 
parameter specifying the attenuation curve assumed. 
For star-forming galaxies, the Calzetti Law \citep{Calzetti2000} is 
widely used. For galaxies with complex stellar populations, 
the two-component dust model of \cite{Charlot2000} may be adopted to 
account for the difference in dust attenuation between star burst clouds 
(stellar populations younger than 10 Myr) and older stellar populations. 
As different attenuation curves are very similar 
to each other in the optical and NIR bands, we 
follow \cite{Charlot2000} and use a single optical depth 
parameter to describe the attenuation of the stellar 
population in a galaxy. 
  
\subsubsection{The implementation of BIGS}
\label{ssec:BIGS}
In complex problems, such as the spectrum fitting problem addressed here, 
the likelihood function can be complicated and may not be represented by 
simple analytical functions. One thus needs an efficient sampling method 
to sample the posterior distribution. In addition, as 
the Bayesian evidence ratio involves integration in high-dimensional 
space, an effective numerical method is also needed to evaluate it.   
BIGS adopts a Bayesian sampler, MULTINEST \citep{Feroz2009,Feroz2013}, 
which uses the nest sampling algorithm to estimate the 
Bayesian evidence, and gives the posterior distribution as a by-product. 

Briefly, BIGS works as follows. For each data spectrum, 
we pre-process it using pPXF \citep{Cappellari2017} and obtain the velocity distribution
of the source. We then convolve our template spectra with a Gaussian 
that accounts for both the instrumental resolution and 
velocity dispersion of stars. The data spectrum and the templates 
are then provided to BIGS. Using the prior distribution of model parameters, 
BIGS uses MULTINEST to generate a proposal parameter vector for the 
spectral synthesis model. These parameters, which specify
the SFH, metallicity and dust attenuation of the stellar population,  
are used to generate a model spectrum to be compared with the 
data spectrum to calculate the corresponding likelihood. The MULTINEST sampler would 
either accept or reject the proposal according to the posterior probability and 
generates a new proposal for the model parameters, until a convergence criterion is 
reached. Once converged, the posterior distribution of model parameters, 
together with the Bayesian evidence, are stored for 
statistical analyses of the model.

\subsection{The fitting procedure}
\label{ssec_fittingproc}

We compare spectra predicted by the E-MILES SPS model 
to the observed MANGA spectra to infer the SFHs of the low-mass galaxies
using the procedure described below. To begin with, we mask some of the 
spectral regions to ensure the validity of the fitting. For example,  
we mask the observed continua in the wavelength range
6800-8100 {\AA}, as difficulties are commonly found in fitting the 
observed continua in this wavelength range, either due to issues in 
flux calibrations in the templates, residual telluric absorption, 
or even flux calibrations in the data 
(see \citealt{Zhou2019} for more discussions). 
In addition, as the E-MILES templates do not contain the youngest 
stellar population (<0.06 Gyr), we also mask the very blue end of the 
spectra (<3800 {\AA}) in the fitting. 

To take into account effects of stellar kinematics and instrumental 
resolution, we first use the software pPXF to pre-fit the data spectra and get an effective velocity dispersion, 
$\sigma_{\rm ppxf}$. This 
effective velocity dispersion is then used to convolve with the E-MILES template 
spectra to generate artificially broadened templates that are used to 
compute the synthesised spectra to be compared with the 
corresponding data spectra.  In this step, apparent emission lines in the spectra 
are also identified, and masked out in subsequent analyses.

After this pre-processing, we first normalise the model and data spectra in 
the wavelength window $4500-5500$ {\AA} and then send them to BIGS. 
BIGS runs the fitting loop as described in \S\ref{ssec:BIGS}, 
assuming a flat prior and a $\chi^2$-like likelihood function. 
In the fitting that uses only the MaNGA spectra, the likelihood 
function is defined as
\begin{equation}
\label{likelyhood}
\ln {L(\theta)}\propto-\frac{1}{2}\sum_{i,j=1}^N\left(f_{\theta,i}-f_{D,i}\right)\left({\cal
M}^{-1}\right)_{ij}\left(f_{\theta,j}-f_{D,j}\right)\,
\end{equation}
where $N$ is the total number of wavelength bins, $f_{\theta}$ and $f_{D}$ are
the flux predicted from the parameter set $\theta$ and that of the data spectrum,
respectively, and ${\cal M}_{ij}\equiv
\langle \delta f_{D, i}\delta f_{D, j} \rangle$
is the covariance matrix of the data.  For spectra that have 
UKIDSS observations, we use the following definition:
\begin{equation}
\label{eq:likelyhood_NIR}
\ln {L(\theta)}\propto-\frac{1}{2}\sum_{i,j=1}^N\left(f_{\theta,i}-f_{D,i}\right)\left({\cal
M}^{-1}\right)_{ij}\left(f_{\theta,j}-f_{D,j}\right)-
\frac{(K_{\theta}-K_{D})^2}{2\sigma_{K}^2}\,
\end{equation}
where $K_{\theta}$ and $K_{D}$ are the $(g-K)$ colour predicted from the 
parameter set $\theta$ and that from the data, respectively. 
The uncertainty in the $(g-K)$ colour is denoted by $\sigma_{K}$. 
In general, it is difficult to model $\sigma_{K}$.
The value of $\sigma_{K}$ is related to the accuracy of UKIDSS photometry, 
which has an uncertainty of less than 2\% in the $K$ 
band \citep{ Dye2006}. In addition, $\sigma_{K}$ should also include 
the relative flux variation between the MaNGA stacked 
spectra and the NIR photometry, which is hard to model. 
To minimize such influence, we only select galaxies with 
$\Delta (g-r)<0.05$ between the MaNGA and UKIDSS archive data. 
In this case, our test shows that setting $\sigma_{K}=0.02$ 
is appropriate to describe the constraints from the NIR observation.
We list all the fitting parameters 
for the $\Gamma$+B model in Table \ref{tab:SFH_parameters}, 
together with their prior distributions (assumed to be flat). 
For the priors of the step-wise SFH, we assume that 
the SFR at the last time interval, 10-14 Gyr, is a constant 
normalized to be 1 (0 in logarithmic scale), and that the prior distributions 
of the SFR in all other time bins are flat between $-2$ and $2$
in logarithmic space. All other parameters are 
specified in the same way as for the $\Gamma$+B model.

\begin{table}
	\centering
	\caption{Priors of model parameters used to fit galaxy spectra}
	\label{tab:SFH_parameters}
	\begin{tabular}{lccr}
		\hline
		Parameter & description & Prior range\\
		\hline
		$\log(Z/Z_{\odot})$ & Metallicity & $[-2.3, 0.2]$\\
		$\tau$ & SFH parameter in Eq.\,(\ref{gamma-sfh}) & $[0.0,10.0]$\\
        $\alpha$ & SFH parameter in Eq.\,(\ref{gamma-sfh}) &$[0.0,20.0]$\\
        $\tau_v$ & Dust optical depth at 5500 \AA &  $[0.0,2.0]$\\
        $f_{\rm burst}$ & relative fraction of old populations & $[0.0,1.0]$\\
        $\log(Z_{burst}/Z_{\odot})$ & Metallicity of the old population  & $[-2.3, 0.2]$\\
        $A_{burst}$ (Gyr) & Age of the old population & $[0.0,14.0]$\\
		\hline
	\end{tabular}
\end{table}

\section{Results}
\label{sec:results}

\subsection{Bayesian model selection}
\label{ssec:modelselection}

\begin{figure}
\includegraphics[height=70mm]{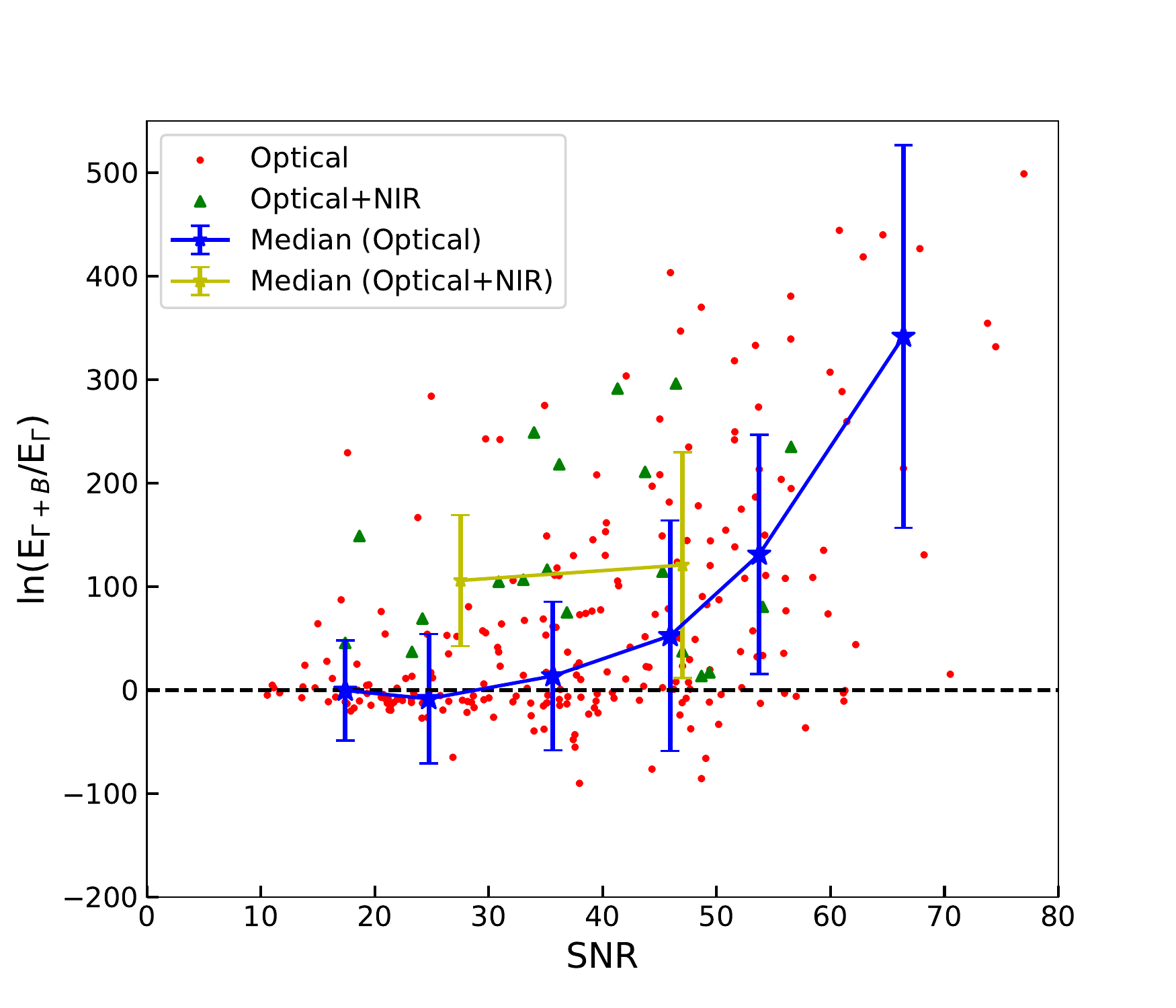}\\\
\caption{The evidence ratio between the $\Gamma$+B and 
$\Gamma$ models as a function of SNR of the stacked spectra. 
Each red dot stands for the result of 
a MaNGA galaxy obtained by fitting its stacked spectrum. Blue stars 
are the median values in five SNR bins and are linked by a blue line. 
Each of green triangles stands for the result obtained from fitting both 
the MaNGA stacked spectrum and the $(g-K)$ (see \ref{ssec_nir}). 
Yellow stars are the median values in two SNR bins and are connected 
by a yellow line. Error bars are $1\sigma$ scatter among galaxies 
in individual bins.
}
\label{fig:evratio}
\end{figure}

\begin{figure}
\centering
\includegraphics[height=65mm]{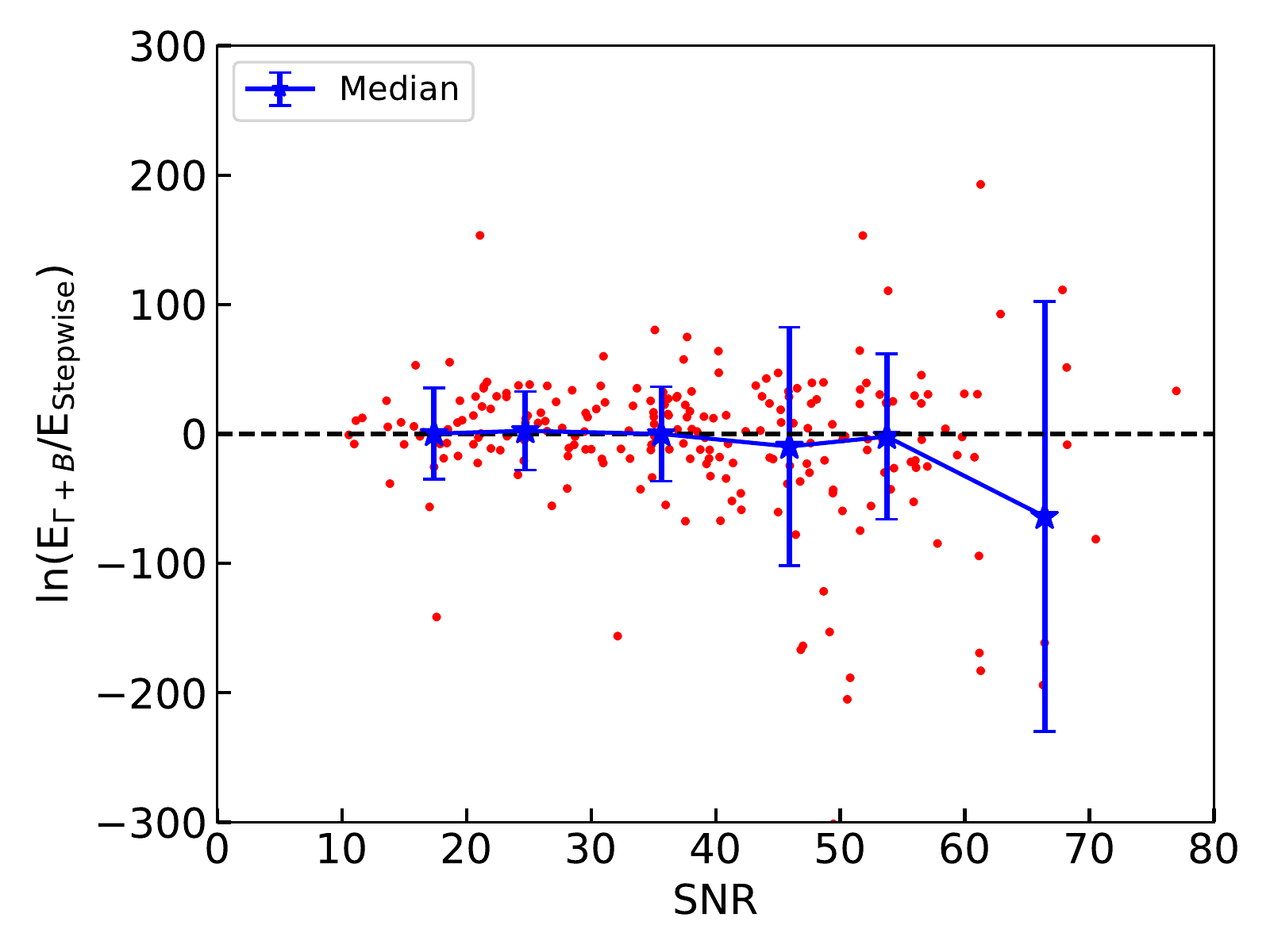}\\\
\caption{The evidence ratio between the $\Gamma$+B and stepwise models 
as a function of SNR of the stacked spectra. Each red dot stands for 
the result of a MaNGA galaxy obtained by fitting its stacked spectrum. 
Blue stars are the median values in five SNR bins and are linked by 
a blue line. Error bars are $1\sigma$ scatter among galaxies 
in individual bins.
}
\label{fig:evratio_step}
\end{figure}

\begin{figure*}
\includegraphics[height=0.35\textwidth]{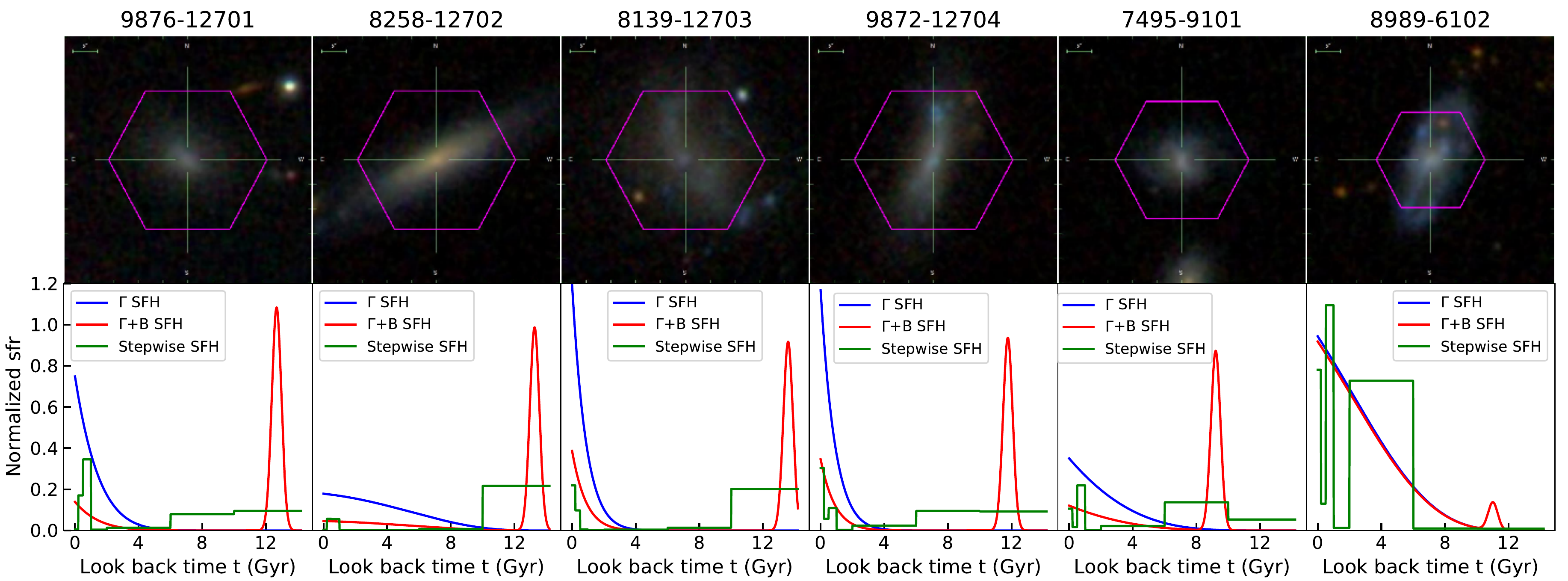}\\\
\caption{
Top panels: optical images of six example galaxies. Bottom panels: the 
corresponding best fit SFHs inferred from the three SFH models, as labelled.
}
\label{fig:SFH_real_example}
\end{figure*}

Our goal is to investigate the existence or absence  
of an old stellar population in low-mass galaxies. 
To this end, we fit the stacked spectra (each being a stack of 
pixels within the effective radius of a galaxy) of individual 
galaxies with three SFH models: $\Gamma$, $\Gamma$+B 
and stepwise, while keeping all other parts of the model intact. 
We first examine if the data shows a preference to one of the three models.
As described above, the Bayesian evidence ratio provided by BIGS 
can serve as a discriminator between different model families. 
Fig.~\ref{fig:evratio} shows the evidence ratio between the two continuous models, $\ln(E_{\rm \Gamma+B}/E_{\rm \Gamma})$, obtained from 
the stacked spectra, as a function of the signal-to-noise ratio 
(SNR) of the stacked spectra. The median values in five SNR bins 
are shown as blue stars to represent the global trend, 
with the error bars showing the $1\sigma$ scatter 
among galaxies in individual bins.

Fig.~\ref{fig:evratio} shows clearly that the median value of the 
evidence ratio increases with increasing SNR.This indicates that the SFHs of these low-mass galaxies are more likely 
a composite of two distinct stellar components than a single 
component. Although the young stellar component may dominate the 
luminosity and makes the galaxies blue, the faint old stellar 
component that formed early may contribute significantly to their 
total stellar masses, as we will quantify next. 

As comparison, we plot in Fig.~\ref{fig:evratio_step} the evidence ratio 
between $\Gamma$+B and the step-wise models. Although the number of free parameters 
in the step-wise model is much larger than that in the $\Gamma$+B model, 
the Bayesian evidences of the two models are comparable. The medians of the evidence ratio are close to 1, with 
large dispersion among different galaxies. This indicates that 
some of the galaxies have a preference to the $\Gamma$+B model, 
while others have the opposite preference. The absent of a systematic 
trend suggests that both models have similar abilities to describe 
the overall SFH in the current data, while the preferences 
of different galaxies to different models indicate the 
intrinsic variations in the detailed shapes of their SFH.

It is, however, not straightforward to quantify the 
preference between different models. Bayesian model selection 
has been used for such purpose, but a quantitative criterion for 
model selection is still not established.
The widely adopted Jeffreys' scale uses a ratio of 
$E_1/E_2>150$, or $\ln(E_1/E_2)>5$, where $E_1$ and $E_2$ are 
the evidences of the two competing models $1$ and $2$, respectively, 
to indicate a 'strong' preference to model 1
\citep[e.g.][]{Hobson2009}. But the validity of this scale 
has been questioned \citep[e.g.][]{Nesseris2013}. 
In spectral fitting, the situation may be even worse. 
The number of data points is large
(thousands of pixels for each spectra) and the current SSP 
models are still not perfect, which may lead to very large 
evidence ratios between different models 
\citep[e.g.][]{Han2019}. In what follows, we simply assume 
that a model is preferred when it has the largest evidence 
among all models considered.

\subsection{The stellar populations in low-mass galaxies}

\subsubsection{The star formation history}

\begin{figure}
\includegraphics[height=0.4\textwidth]{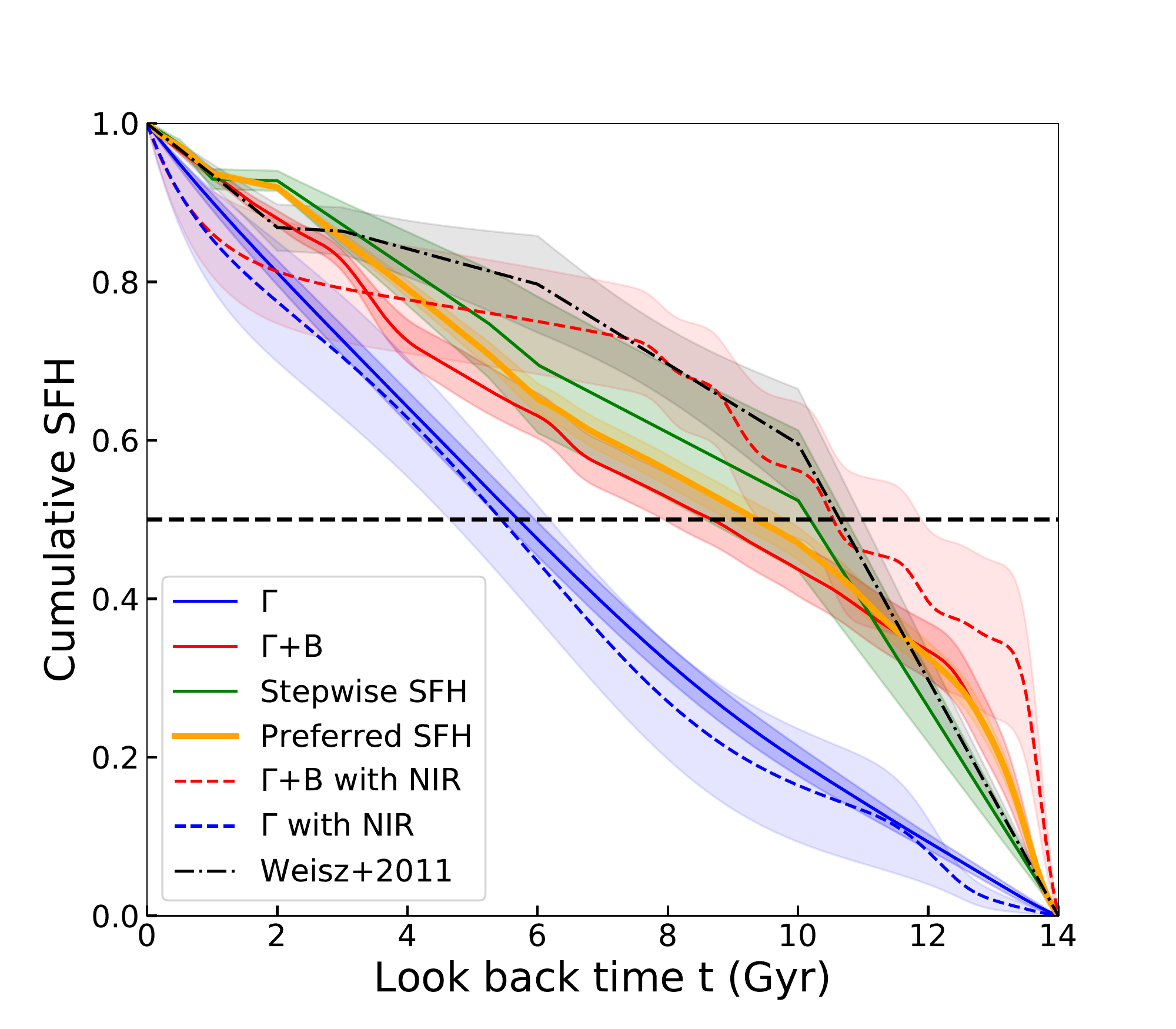}\\\
\caption{
The cumulative SFH inferred from the posterior distribution 
constrained by the stacked spectra of different galaxies 
within $1 R_e$. The red, blue and green solid lines show the results 
from the best-fit $\Gamma$+B, $\Gamma$ and stepwise models, 
respectively. The thick orange line shows the 
average SFH of the preferred model for each galaxy, 
selected by Bayesian evidence. 
The shaded region around each line represents the variance 
of the mean SFH, estimated from the jackknife 
resampling method. The result of \citet{Weisz2011} is shown as 
a black dash-dotted line.
Results obtained from the MaNGA stacked spectra plus the $(g-K)$ 
colour (see \ref{ssec_nir}) using the $\Gamma$+B model and the $\Gamma$ model are shown with red and blue dash lines, respectively. The horizontal black dashed line marks the position 
of half of the total star formation.
}
\label{fig:cum_SFH}
\end{figure}

\begin{figure}
\includegraphics[height=0.4\textwidth]{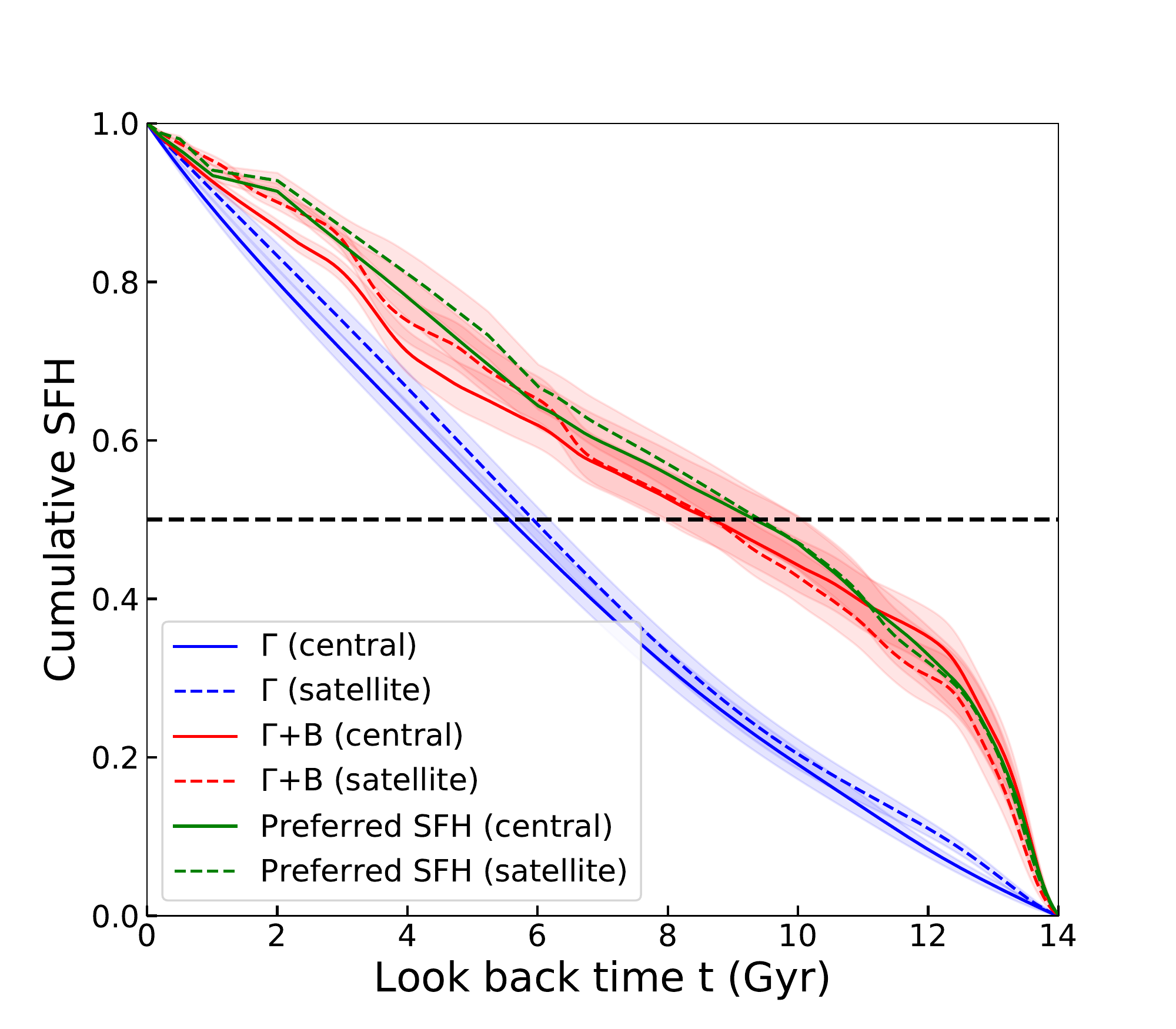}\\\
\caption{
The cumulative SFH inferred from the posterior distribution 
constrained by the stacked spectra of different galaxies 
within $1 R_e$. The red, blue and green lines show the best fits  
of the $\Gamma$+B, $\Gamma$ and preferred SFHs, respectively.
Solid and dash lines are 
mean results for central and satellite galaxies,
respectively. The shaded region around each line represents 
the variance of the mean SFH, estimated from the jackknife 
resampling method. The horizontal black dashed line marks the 
position of half of the total star formation. }
\label{fig:cum_SFH_cs}
\end{figure}
\begin{figure}
\centering
\includegraphics[height=0.4\textwidth]{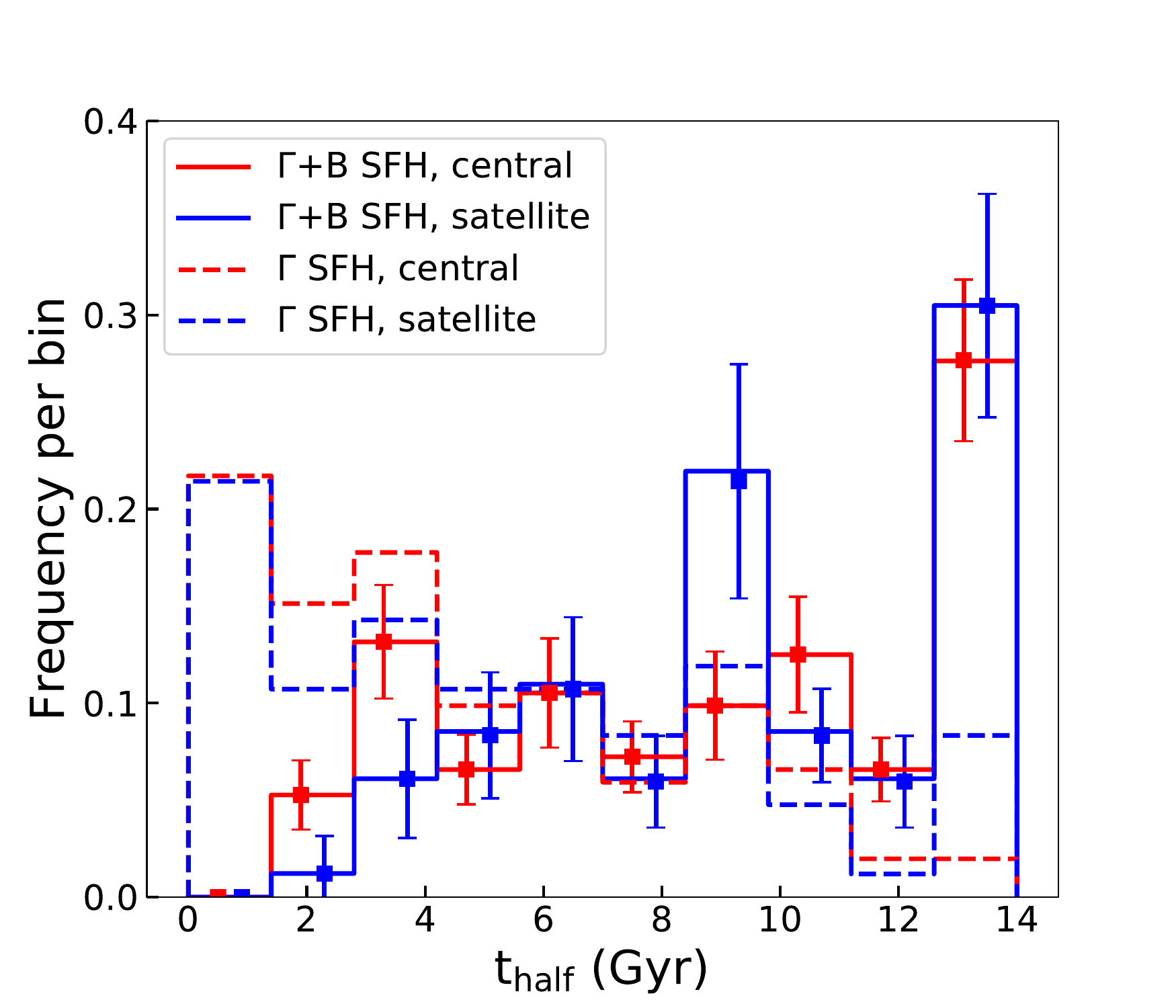}
\includegraphics[height=0.4\textwidth]{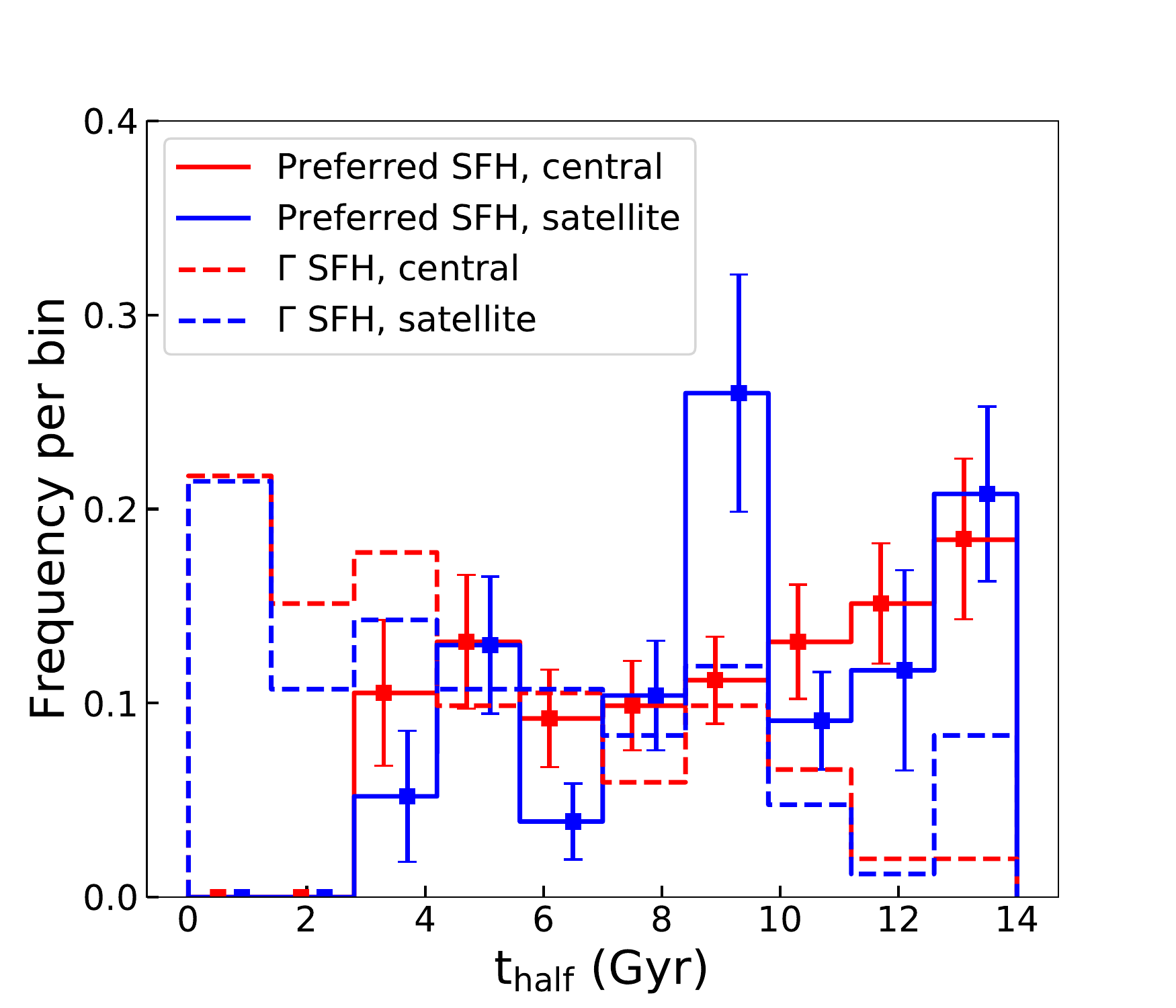}
\caption{The distribution of the half mass formation time inferred from the 
best-fit SFH models. Red and blue histograms show the results of 
central and satellite galaxies, respectively. Solid histograms are obtained using the $\Gamma$+B SFH 
(top panel) and the preferred SFH (bottom panel), 
while dashed ones are for the $\Gamma$ SFH. 
Error bars, only shown for the $\Gamma$+B SFH (top panel)
and the preferred SFH (bottom panel),  are obtained 
from the jackknife resampling method. Each histogram is normalized to 1
over all the bins used.
}
\label{fig:t_half_stack}
\end{figure}

\begin{figure*}
\centering
\includegraphics[height=0.85\textwidth]{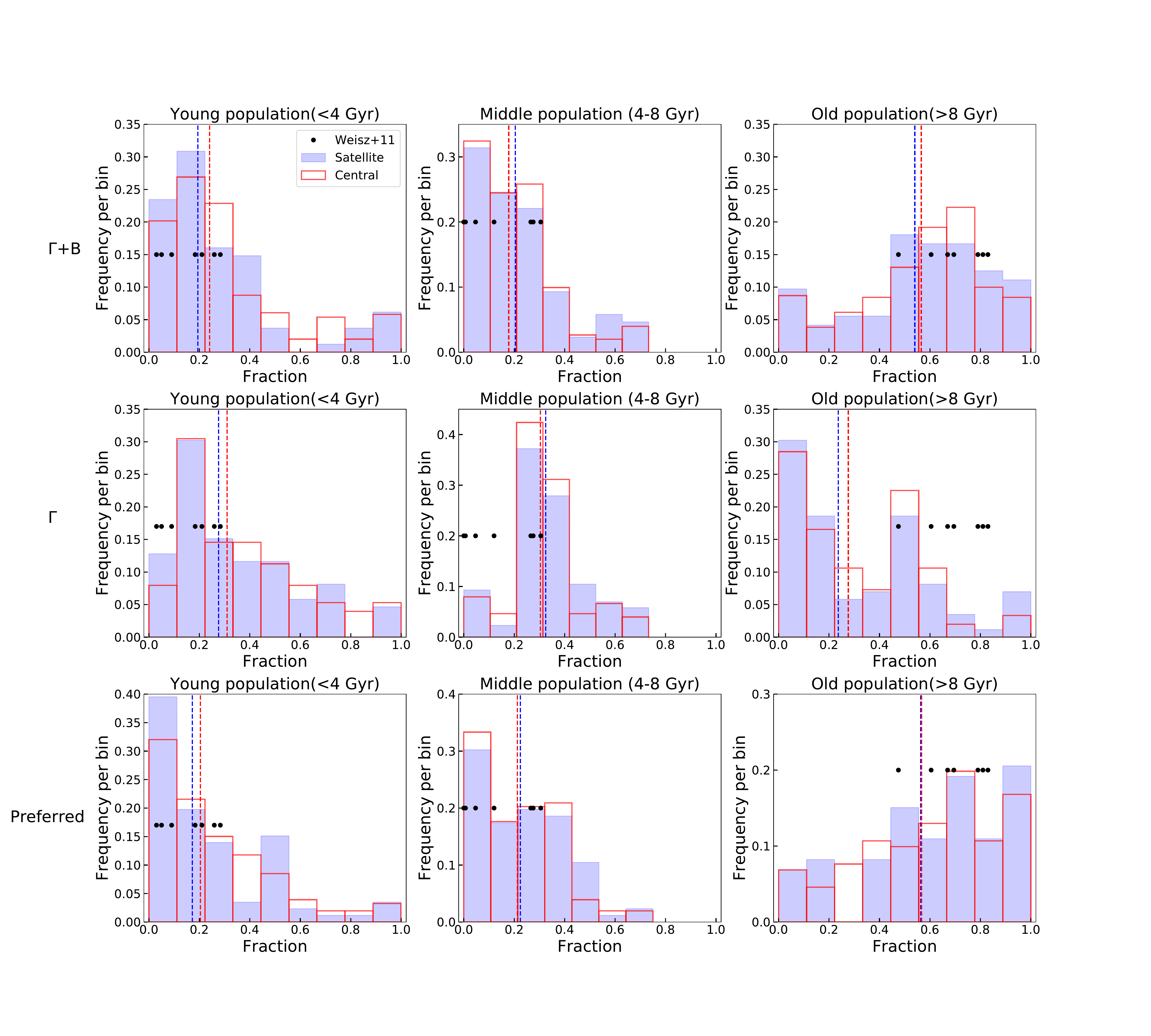}
\caption{The distribution of the mass fraction of stars in the old (>8 Gyr, left), 
middle-age (4-8 Gyr, middle), and young (<4 Gyr,right) stellar populations. Top panels are results obtained from the $\Gamma$+B SFH model, middle panels 
are from the $\Gamma$ model, while bottom panels are from the preferred SFH. Red and blue histograms are 
for centrals and satellites, respectively, 
with dash lines showing the median values. Black dots are results 
of \citet{Weisz2011}. All histograms are normalized to 1
over the bins used.}
\label{fig:hist_fractions}
\end{figure*}

We first obtain the best-fit SFH for each galaxy from 
the posterior distribution. Results for six representative galaxies are shown in 
Fig.~\ref{fig:SFH_real_example}. As can be seen, 
results from the $\Gamma$+B model and the stepwise model 
are broadly consistent with each other, both predicting the 
existence of an old stellar population for most (but not all) 
galaxies, while the $\Gamma$ model clearly misses such a 
population. 

We plot the average cumulative SFH in Fig.~\ref{fig:cum_SFH}. 
To estimate the statistical uncertainty, 
we divide the galaxy sample into 20 sub-samples of 
similar sizes, and make 20 different jackknife copies 
from these sub-samples by eliminating one of the 
sub-samples. The variances of the mean SFH inferred from  
the 20 jackknife copies are shown in Fig.~\ref{fig:cum_SFH} as 
the shaded regions around the corresponding lines.
The results derived from the $\Gamma$+B model (red lines) 
indicate that low-mass galaxies on average formed about  
half of their stellar masses more than 8 Gyrs ago, 
which may be associated with the starburst events observed in 
extreme emission line galaxies detected in the CANDELS 
survey \citep{vanderWell2011}. As a comparison, the black line
shows the cumulative SFHs of local dwarf galaxies in the mass range 
$10^8-10^9{\rm M}_{\odot}$ obtained from resolved 
stars by \citet{Weisz2011}. It is remarkable 
that the results obtained from the $\Gamma$+B model are in 
good agreement with that of \citet{Weisz2011}.  
In contrast, the results obtained from the 
$\Gamma$ model (blue lines) indicate that most of 
the stellar mass in low-mass galaxies formed recently.
This discrepancy is expected. Since an old stellar 
population is much fainter than the young population 
of the same mass, a model that is not sufficiently 
flexible to allow for the existence of both populations 
will miss the old population. As shown in \S \ref{ssec:modelselection}, 
this limitation of the $\Gamma$ model weakens its 
ability to describe the true SFH, which leads to 
the smaller Bayesian evidence in comparison to the 
$\Gamma$+B model in fitting spectra of sufficiently high SNR. 
For comparison, we also show in Fig.~\ref{fig:cum_SFH} the 
result obtained from the stepwise model as a green line. 
In addition, we also obtain the best-fit SFH from the preferred
model for each galaxy based on Bayesian model selection. 
The result, referred to as the {\it preferred SFH}, is plotted 
in Fig.~\ref{fig:cum_SFH} as a thick orange line. 
The two results are in qualitative agreement with those from 
the $\Gamma$+B model, indicating
that the $\Gamma$+B model may be sufficiently 
broad for the current data, and that our conclusion does not 
depend on the exact form assumed for the SFH,  
as long as it is sufficiently flexible.

In what follows, we will thus focus on the 
results derived from the $\Gamma$+B model and use them to 
compare with those in the literature. In addition, we will also
show results from the preferred SFH to demonstrate how our
results may vary due to the use of a different SFH model.

In Fig.~\ref{fig:cum_SFH_cs}, we show results separately for central 
and satellite galaxies, using the central/satellite classification
from Galaxy Environment for MaNGA Value 
Added Catalog (GEMA-VAC, \citealt{Argudo2015}). This VAC uses the 
group catalogue of \cite{Yang2007} to separate MaNGA galaxies 
into centrals and satellites. Environmental effects can be 
seen from Fig.~\ref{fig:cum_SFH_cs}, 
in that satellite galaxies appear to form their stars 
slightly earlier than central galaxies.
This is expected, as low-mass satellites may have their star 
formation quenched by environmental effects of their host 
halos \citep[e.g.][]{vandenBosch2008,Peng2012}.  

In addition to the cumulative SFH, we also derive 
the half-mass formation time, $t_{\rm half}$, defined as the 
look-back time when a galaxy forms half of its final stellar mass, 
from the best-fit SFH models and plot the results in 
Fig.~\ref{fig:t_half_stack}. Again, GEMA-VAC is used to divide 
our sample into centrals (red) and satellites (blue).  
For both the $\rm \Gamma+B$ model and the preferred SFH, 
$t_{\rm half}$ varies from 2 Gyr to 12 Gyr, with a broad peak at 
about 9 Gyr. In contrast, $t_{\rm half}$ 
inferred from the $\rm \Gamma$ model is 
smaller on average, ranging from zero to 12 Gyr with a broad
peak at $<4$ Gyr. 

Comparing results obtained for centrals and satellites, 
one sees that the peak at 9 Gyr is weaker for centrals, and there is 
a weak excess of central galaxies at $t_{\rm half}\sim 4 {\rm Gyr}$. 
This excess may indicate a secondary 
star formation episode for some of the central galaxies,
as is expected from the "gappy" star formation history found 
by \citet{Wright2019}. These results are in rough agreement 
with those of \citet{Kauffmann2014}, who found that the distribution of 
$t_{\rm half}$ for low-mass galaxies, derived from analysis of 
Dn4000 and ${\rm H}_{\delta_A}$ absorption features, 
is quite broad and shows double peaks.

We note that there is a significant peak in the distribution 
of $t_{\rm half}$ at around 13 Gyr inferred from $\Gamma$+B.
However, our extensive test shows that this peak may not 
be real. We only use a single SSP to describe the burst in the 
early-Universe, and the age of the burst is confined to be within the age of 
Universe. As the spectra of old SSPs are insensitive to the age
\citep[e.g.][]{BC03}, a galaxy with an early starburst that contributes 
more than half of its stellar mass could have the best-fit 
$t_{\rm half}$ at the edge of the prior. By setting the prior  
age range to be the maximum age of the SSP model, which is 
18 Gyr for the EMILES model, we found that some best-fit ages 
would move out of the previous boundary, and the peak at 13 Gyr 
would disappear. These results indicate that the current model 
is not able to describe the ages of old stellar populations 
accurately. Because of this limitation, the exact ages of the old 
population predicted by the model should be treated with caution.
We emphasise, however, that this limitation does not affect the 
conclusion that these galaxies contain large fractions 
of old stars. Indeed, the peak is much weaker in the distribution 
inferred from the preferred SFH, but the predicted fraction of 
galaxies with $t_{\rm half}> 8\,{\rm Gyr}$ is similar 
for both the $\Gamma$+B and the preferred SFH.

\begin{figure*}
\centering
\includegraphics[height=0.3\textwidth]{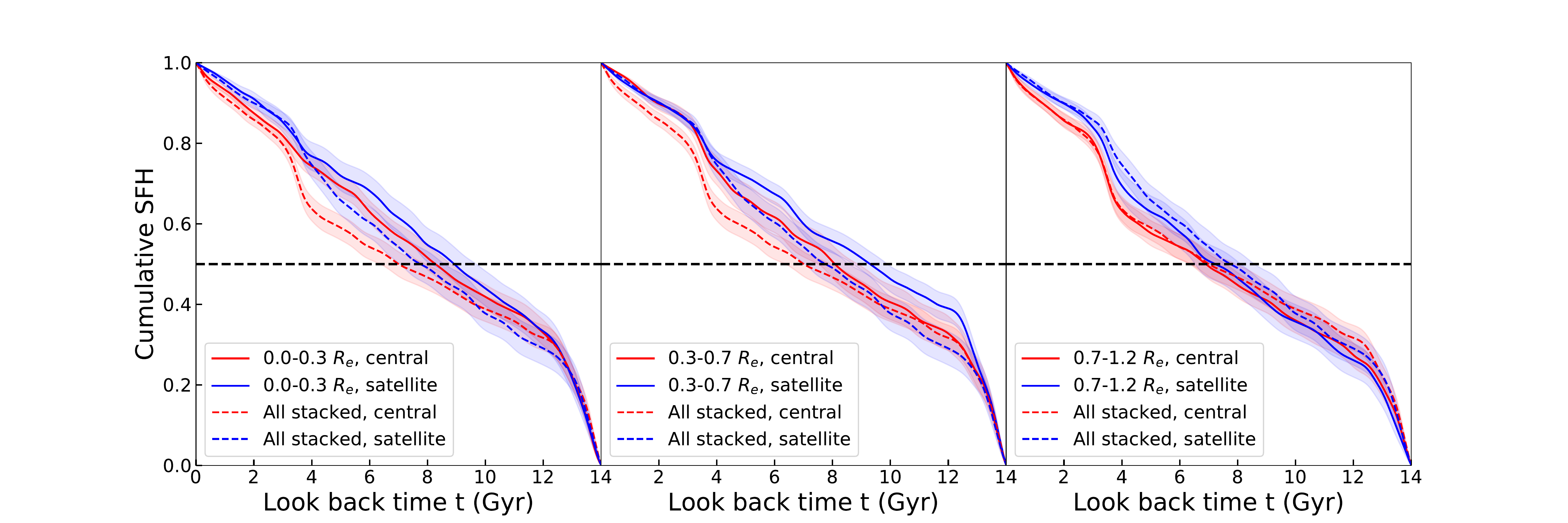}
\includegraphics[height=0.3\textwidth]{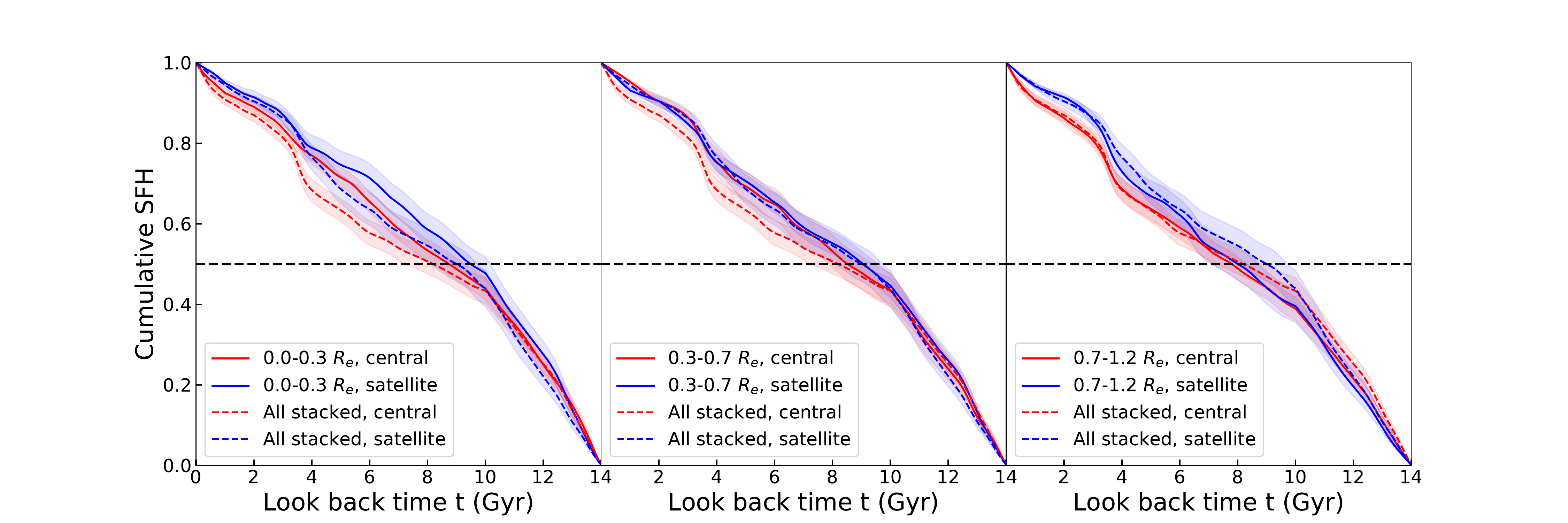}
\caption{The cumulative SFH inferred from the best-fit 
$\Gamma$+B model (top panels) and the preferred SFH 
(bottom panels). Solid lines in all three panels show results for radial bins 
[0.0-0.3]$R_e$, [0.3-0.7]$R_e$, and [0.7-1.2]$R_e$, respectively. 
As a reference, results of stacking spaxels of the entire galaxy are shown 
by dash lines in all panels. Red and blue lines are for central 
and satellite galaxies, respectively. The shaded region around 
each line represents the variance of the mean SFH, 
estimated from the jackknife resampling method. The horizontal black dashed line 
marks the position of half of the total star formation.}
\label{fig:sfh_cum}
\end{figure*} 

\subsubsection{The old stellar populations}

Fig.~\ref{fig:hist_fractions} shows the mass fraction of stars in 
three populations, old (>8~Gyr), middle-age (4-8 Gyr), and young 
(<4 Gyr). We show the results for central and satellites galaxies
separately, and separately for ${\rm \Gamma+B}$ (upper panels),
$\rm \Gamma$ (middle panels) and the preferred SFH (bottom panels).
For ${\rm \Gamma+B}$ and the preferred SFH, 
the fractions are $\sim60\%$, $20\%$ and $20\%$ for the three 
populations, respectively, and are quite similar for both 
centrals and satellites.
For the $\Gamma$ model, the corresponding 
fraction is about one third for each of the three populations. 

Therefore, the $\Gamma$ model significantly underestimates 
the fraction of old stars while overestimating the young
fraction in comparison with the ${\rm \Gamma +B}$ model 
and the preferred SFH. 
For comparison, we show the results obtained from the resolved observation 
of \citet{Weisz2011} as black dots. Although the resolved data is sparse, 
there is a good match of the data with the distribution inferred from 
both $\Gamma$+B and the preferred SFH, 
but a significant mismatch with the inferences of 
the $\Gamma$ model, indicating again that the inclusion of an early episode 
of star formation is the minimal requirement to explain 
the stellar population in low-mass galaxies. 
Note that the amount of stars in the middle age 
(4-8 Gyr) is relatively small both in the predictions of 
$\Gamma$+B/preferred SFH and in the results based on resolved stars.  
This is not well reproduced in hydrodynamic simulations 
\citep{Digby2019}, suggesting that the observational data 
can provide important constraints on the formation and evolution 
of low-mass galaxies. 

\subsubsection{Radial dependence}
\label{ssec_radial}

The spatially resolved spectra provided by MaNGA allow us to investigate 
the SFHs in different parts of a galaxy. Here we use MaNGA spectra stacked 
in four radial bins, $[0.0-0.3)R_e$, $[0.3-0.7) R_e$, $[0.7-1.2)R_e$, 
and the entire galaxy with all spaxels, 
to investigate how the SFH varies with radius. 
We derive the average SFH from the radially stacked spectra 
using the $\Gamma$+B model 
and the step-wise model. Fig.~\ref{fig:sfh_cum}
shows the cumulative distribution of the SFH obtained for the four
radial intervals. Results are shown for the $\Gamma$+B model 
(top panels) and the preferred SFH (bottom panels), 
and separately for central and satellite galaxies. 
It is seen that the old stellar population exists not 
only in the central part, but spreads over the entire galaxy, 
making a significant contribution to the total stellar mass.
Satellites form their stars slightly earlier than
centrals and this is true for all radii. In addition, there 
is an indication that stars in the innermost part on 
average formed earlier, by $\sim 1\,{\rm Gyr}$, than 
in the outer part. However, the signal for the age 
gradient is too weak and the uncertainty in the result is 
too large to draw a definite conclusion.

\subsection{Constraints from NIR}
\label{ssec_nir}

\begin{figure*}
\centering
\includegraphics[height=0.4\textwidth]{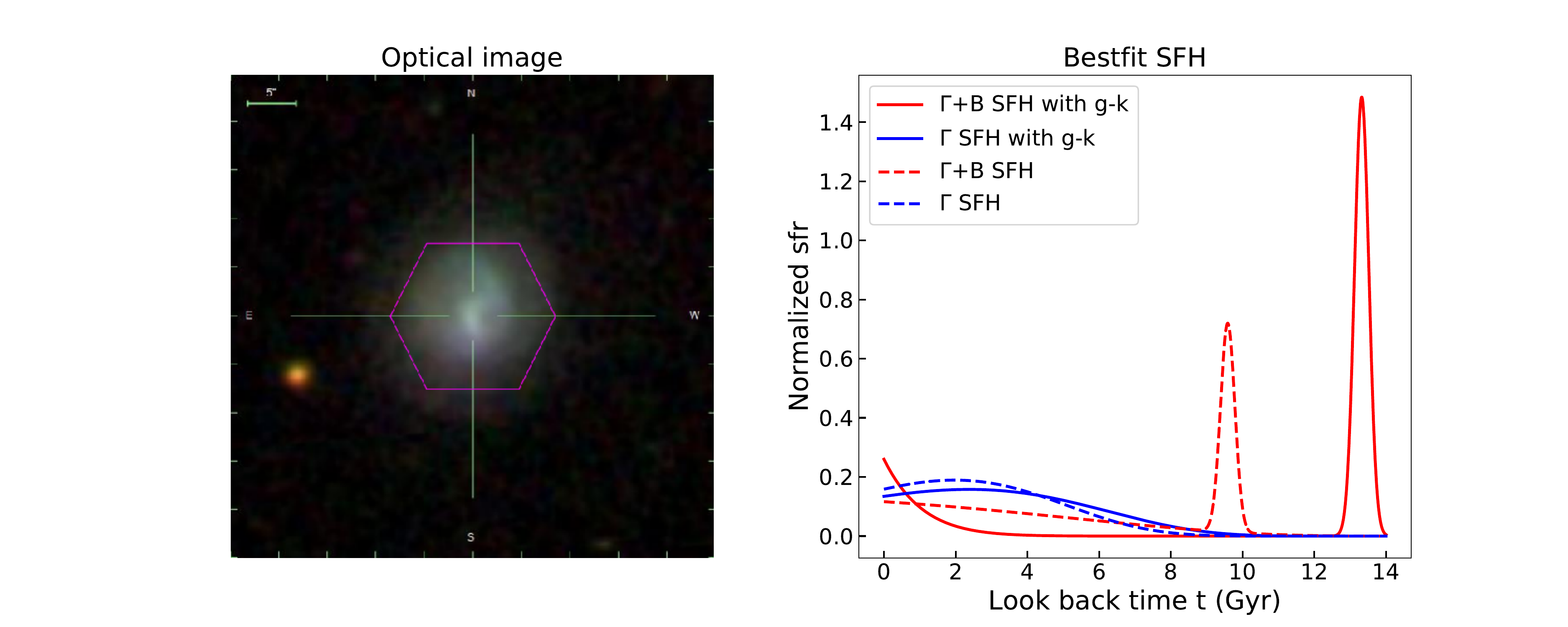}
\caption{ An example showing the changes in the inferred SFH by including 
UKIDSS NIR colour in the fitting. The left panel shows the optical 
image of this galaxy, and the right panel shows the best fit SFHs 
from different models. Red and blue lines show the results from the 
$\Gamma$+B and $\Gamma$ models, respectively. Dash lines are 
results from the stack spectra from MaNGA, while solid lines are 
results that include the $(g-K)$ colours.}
\label{fig:SFH_nir}
\end{figure*}

The results presented above are obtained using MANGA optical 
spectra as constraints. The SSP templates from the E-MILES model 
in fact have wavelength coverage from 1600 {\AA} to 5 ${\rm \mu m}$. 
Thus, these SSP templates allow us to predict the fluxes in
both optical and NIR bands once a set of model parameters are given.
By comparing the predicted colour with observations, we can get additional
constraints on the stellar population of galaxies.   
Compared to spectroscopic observations, broad band measurements 
are much easier to make, although they may lose some spectral information.
As old stars emit the majority of their light in NIR, 
a combination of optical spectra and 
NIR photometry is expected to provide better constraints on
the SFH model which includes early star formation.
In this subsection, we demonstrate this 
using the combination of UKIDSS $K$-band flux and MaNGA optical 
spectrum.

The fitting example of mock spectra in Fig.~\ref{fig:SFH_example} 
illustrates the improvement of the derived SFH when $(g-K)$ colour 
is included in the fitting. As a real example, Fig.~\ref{fig:SFH_nir}
shows the fitting results for MaNGA 9876-3703, one of the galaxies
in our sample.   
From the optical image shown in the left panel, one can see that 
this galaxy is quite blue, indicative of ongoing star formation.
Indeed, if we assume a simple $\Gamma$ model to fit
its MaNGA spectra, the SFH obtained is dominated by recent star formation, 
with almost no population older than 8 Gyr, as shown by the blue lines. 
In contrast, the use of the $\Gamma$+B model reveals the existence of 
a significant old stellar population. However, neither of 
the two constrained models can recover perfectly the observed 
$(g-K)$ colour for this galaxy. Using the posterior distribution of 
the model parameters obtained from the E-MILES templates, 
the best-fit of the $\Gamma$ and $\Gamma$+B models predicts 
$(g-K)=2.08$ and $(g-K)=2.22$, respectively, while the measurement 
from WSA is $(g-K)=2.43$. These results indicate that 
the lack of constraints from NIR may still lead to biased inferences 
of the stellar population, even if a proper SFH is assumed. 

In order to check the effects of including the NIR photometry, 
we implement a new set of fitting, adopting the modified likelihood 
function described by equation (\ref{eq:likelyhood_NIR}). 
The results obtained for MaNGA 9876-3703 are plotted in 
Fig.~\ref{fig:SFH_nir} with solid lines. The inclusion of 
the $(g-K)$ colour changes the fitting result of  
the $\Gamma$+B model significantly, in that the fraction
of the old population increases significantly.
The predicted $(g-K)$ colour is $2.38$, very close to the 
observed value. In contrast, the result obtained from the $\Gamma$ model 
is not affected as much by the inclusion of the NIR data, 
with $(g-K)=2.27$, which is still too blue.

We apply the same fitting to all the 19 galaxies that have UKIDSS NIR 
photometry. The green triangles in Fig.~\ref{fig:evratio} show the 
evidence ratio between the $\Gamma$+B and the $\Gamma$ models 
as a function of SNR of the stacked spectrum. Yellow stars
are the median of the values in two SNR bins, divided at $SNR=40$.
Compared to the red dots that only use the MaNGA stacked 
spectra, there is a significant increase in the evidence ratio, indicating 
an increase in the ability of discriminating the two SFHs.

The red and blue dash lines in Fig.~\ref{fig:cum_SFH} 
show the cumulative SFHs for the NIR sample obtained from the 
best-fit $\Gamma$+B and $\Gamma$ model, respectively. 
Compared to the solid lines that show results using MaNGA spectra 
alone, the result obtained with the $\Gamma$+B model including 
the NIR data shows a significant increase in the stellar mass
formed 8 Gyr ago, although the uncertainty is large  
owing to the limited sample size. In contrast, the result
obtained by the $\Gamma$ model does not change much.

\begin{figure}
\centering
\includegraphics[height=0.4\textwidth]{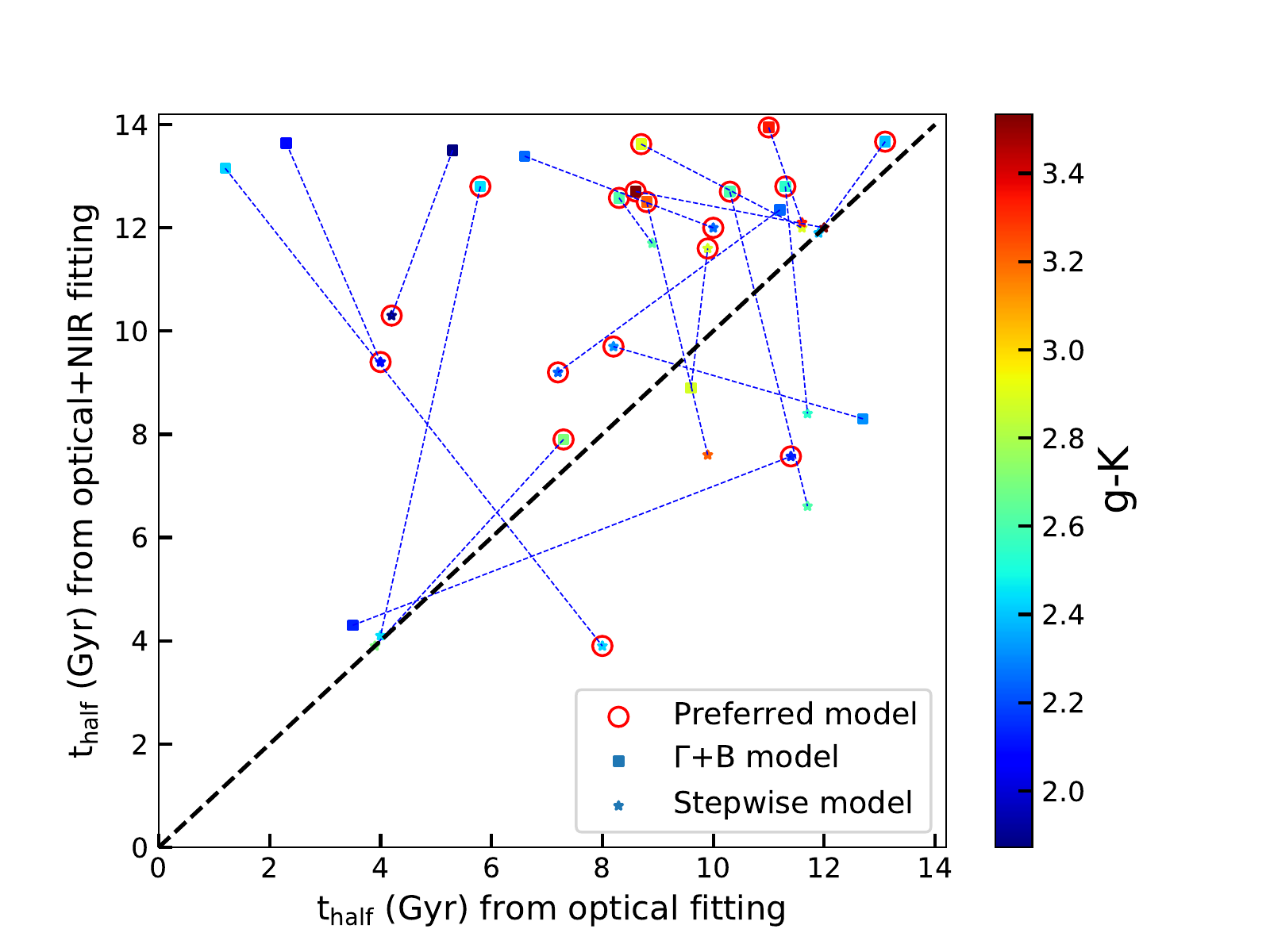}
\includegraphics[height=0.4\textwidth]{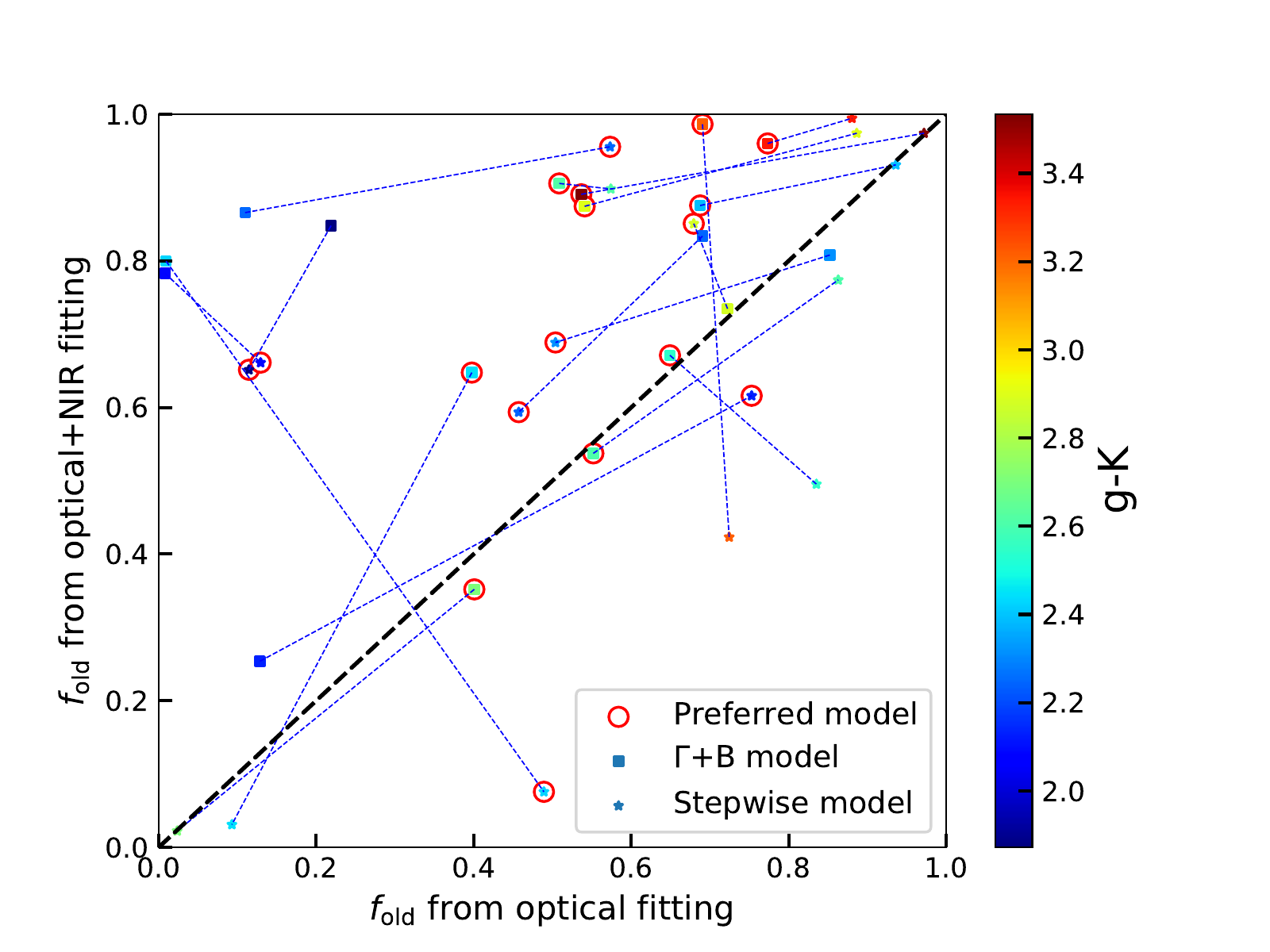}
\caption{Comparison of the half mass formation time (top) and the fraction 
of old population (bottom) between fitting only the MaNGA stacked 
spectra (horizontal axis) and fitting both optical spectra and 
NIR photometry (vertical axis), colour coded by the $(g-K)$ 
colour of galaxies. Squares are results obtained from the $\Gamma$+B model, while stars are results from the stepwise model. 
Results from the same galaxy are linked by a blue dash line, 
while the preferred SFH model for each galaxy is marked by 
a red circle.}
\label{fig:compare_nir}
\end{figure}

In Fig.~\ref{fig:compare_nir}, we compare the half mass formation 
times and the fractions of old (>8 Gyr) population derived from 
optical-only and optical plus NIR data. Results are shown for both the $\Gamma$+B model (squares) and the stepwise model (stars). We 
also indicate the preferred SFH with a circle. 
It is seen that for most galaxies, the inclusion of 
the NIR constraint tends to increase $t_{\rm half}$, 
or equivalently, to produce a higher fraction 
in the old population. This suggests that the 
old fraction inferred from the optical spectra 
alone may be an underestimate.

However, we should point
out that those results also raise a concern. 
The tension between the predictions based on optical spectra
and NIR photometry indicates that the model adopted may not be 
sufficiently general. To examine this problem in more detail,  
we apply the posterior predictive check method to the sample
of 19 galaxies with NIR photometry. The detail is presented in the 
Appendix \ref{appedix_nir}. 
In general, we find that the model is often over-constrained
by the optical spectra, so that the posterior predictive distribution 
(PPD) of the NIR photometry is very narrow and the observed NIR data
is almost always rejected by the posterior predictive check (PPC).
In Bayesian statistics, the inferences obtained from a data set 
apply only to the model (hypothesis) assumed. If the model is 
not general enough to accommodate all the information in the data, 
inferences may still be made for the assumed model. 
In this case, one might want to use all available data, 
hopefully to obtain a balance between different constraints. 

In summary, the combination of the NIR photometry and optical 
spectra provide additional evidence that an early old stellar population 
exists in most of the low-mass galaxies. However, due to the limitation 
of the assumed model, the current analysis is unable to reach a 
quantitative consistency between the optical-only and 
optical$+$NIR results. This should be kept in mind when 
one is concerned with the details of the inferred SFH.

\begin{figure*}
\centering
\includegraphics[height=70mm]{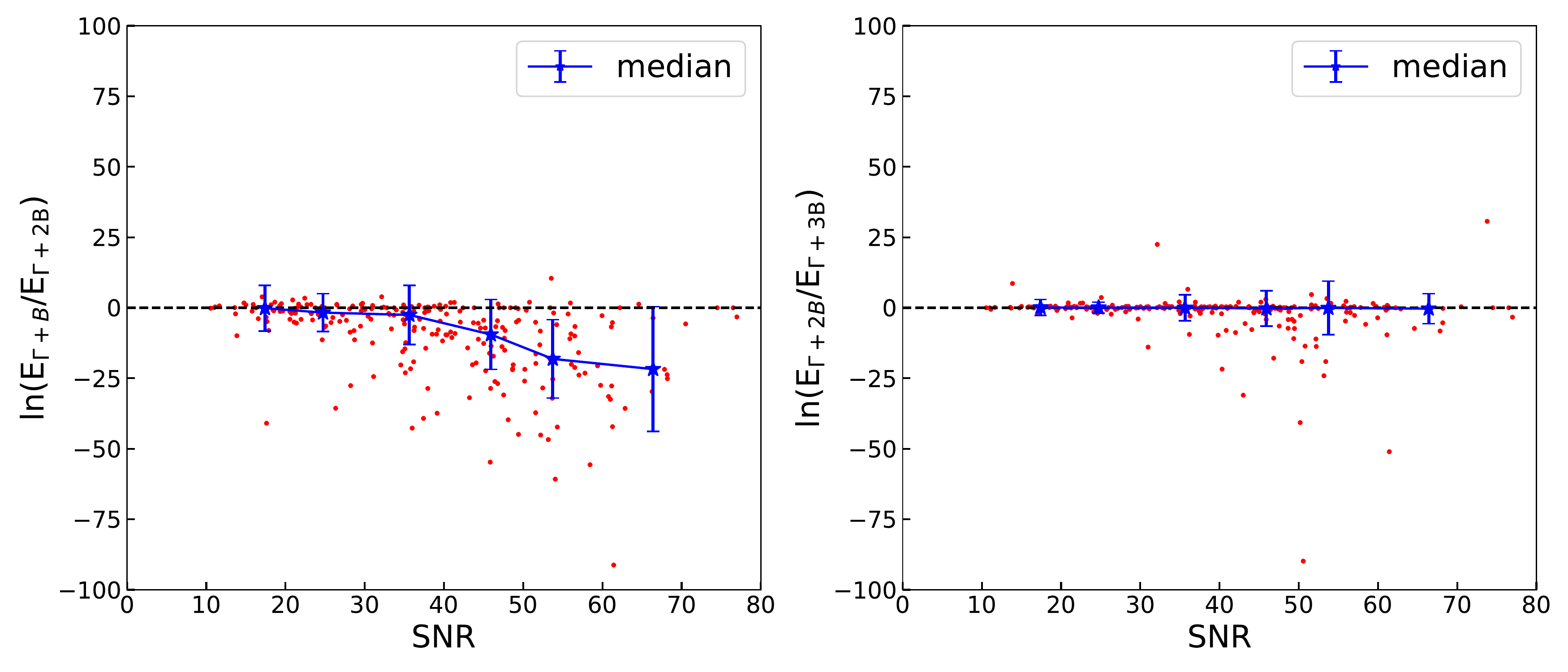}
\caption{
The evidence ratio between the $\Gamma$+B and $\Gamma$+2B models (left panel) 
and between the $\Gamma$+2B and $\Gamma$+3B models (right panel) 
as a function of SNR of the stacked spectra.Each red dot stands for 
the result of a MaNGA galaxy. Blue stars connected by a blue line
are the median values in five SNR bins. Error bars are $1\sigma$ scatter among galaxies in individual 
bins.
}
\label{fig:sfh_compare_scburst}
\end{figure*}

\begin{figure}
\centering
\includegraphics[height=0.7\textwidth]{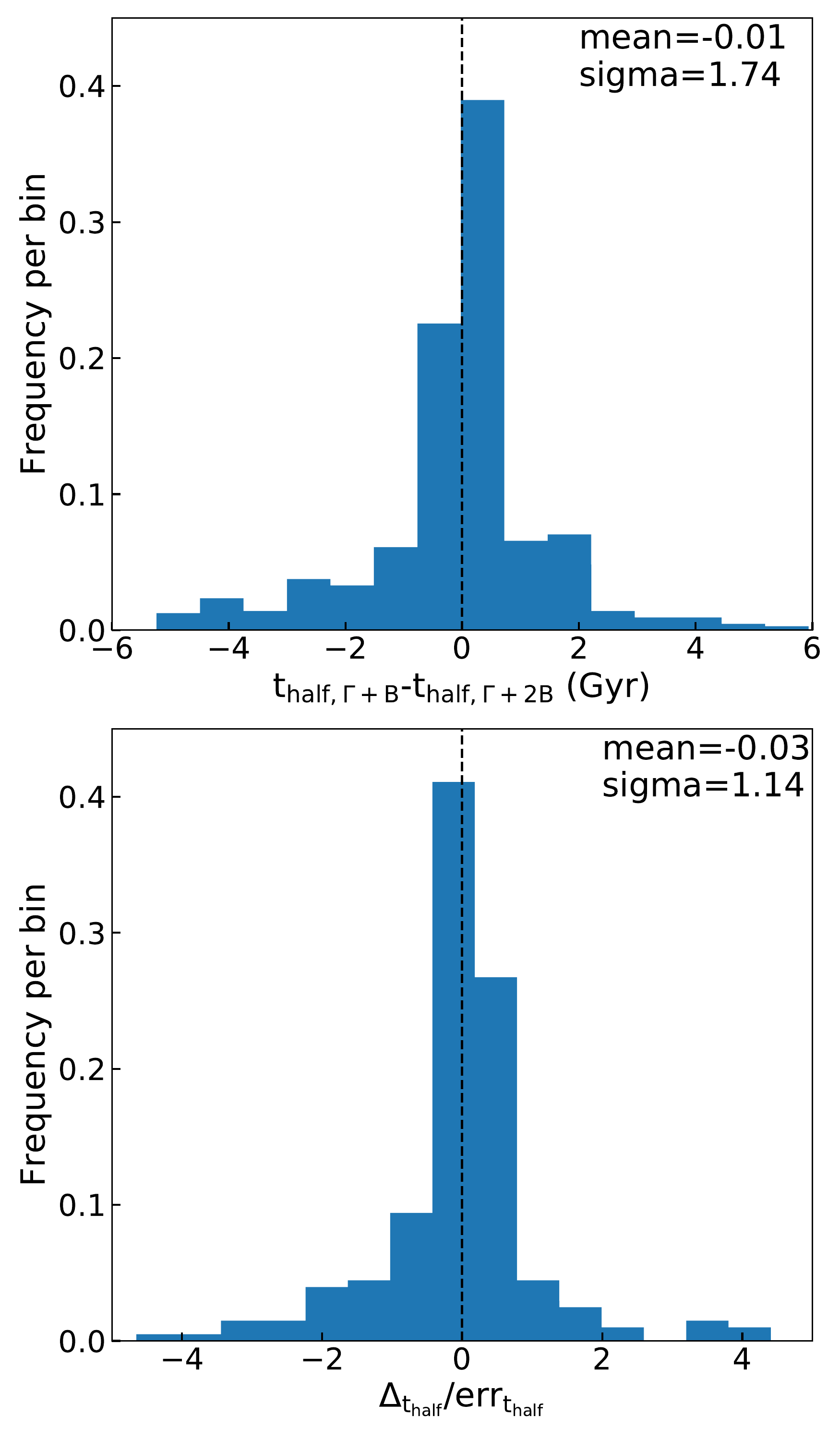}
\caption{
The distribution of the difference in the half mass formation time 
between the $\Gamma$+B and $\Gamma$+2B models (top), and 
the distribution of the difference normalized by the 
inference uncertainty (bottom). The mean value and standard deviation 
are indicated for the histogram in each panel.
}
\label{fig:sfh_compare_scburst_thalf}
\end{figure}

\section{Uncertainties}
\label{sec:Uncertainties}

The results obtained above are based on the analysis of 
MaNGA spectra and UKIDSS NIR photometry with the use of BIGS. 
We adopt the state-of-the art SPS model based on the E-MILES 
SSP templates, and assume three models for the SFH.   
In this section we examine further potential uncertainties 
in different parts of our analysis.
 
\subsection{SFH models}

 We first examine whether or not our SFH model is sufficient to 
characterise the SFH of real galaxies.
To this end, we first extend our SFH model by including additional 
bursts in the fitting. Models that have two and three bursts are  
referred to as the $\Gamma$+2B model and $\Gamma$+3B model, respectively. Fig.~\ref{fig:sfh_compare_scburst} shows the evidence ratio 
between these models as a function of SNR. Compared with 
$\Gamma$+B model, the $\Gamma$+2B model is preferred by 
some galaxies, but the global trend for this preference is mild. 
The evidence ratio between $\Gamma$+2B and $\Gamma$+B is much smaller than 
that between $\Gamma$+B and $\Gamma$ shown in Fig.~\ref{fig:evratio}, 
indicating a weaker preference for more complex models. 
Inspecting The evidence ratio between  
$\Gamma$+2B and $\Gamma$+3B model, one can see that, 
even for galaxies which prefer a second burst, a third burst 
is unnecessary. These results show again that, given the current 
data quality and the SPS model, the $\Gamma$+B model seems 
to be sufficient.

We further test whether the additional burst changes our 
basic conclusion. To this end, we estimate the difference between 
$t_{\rm half}$ derived from SFH models with one and two bursts, 
denoted by $\Delta t_{\rm half}$, for individual galaxies. 
In addition, we compare $\Delta t_{\rm half}$ with the 
inference uncertainty: 
${\rm err}_{ t_{\rm half}} \equiv 
\left(\sigma^2_{\rm t_{\rm half,~\Gamma +B}} + 
\sigma^2_{\rm t_{\rm half,~\Gamma + 2B}}\right)^{0.5}$, 
where $\sigma_{\rm t_{\rm half,~\Gamma +B}}$ and 
$\sigma_{\rm t_{\rm half,~\Gamma +B}}$ are the standard deviations 
of the posterior distribution of $t_{\rm half}$, 
inferred from the $\Gamma$+B and $\Gamma$+2B models, respectively.  
We use the ratio between $\Delta t_{\rm half}$ and ${\rm err}_{ t_{\rm half}}$ 
to characterize how well the differences in the derived 
$t_{\rm half}$ are accounted for by the inference uncertainty. 
The results plotted in Fig.~\ref{fig:sfh_compare_scburst_thalf}
show that adding a new burst into the model have almost negligible 
effects on the derived $t_{\rm half}$. The differences between the 
derived $t_{\rm half}$ are only slightly larger than the 
inference uncertainty, indicating that the $\Gamma$+B 
model is an acceptable choice for our purpose.

We also briefly discuss the differences between the 
parametric and non-parametric models. Non-parametric models avoid 
the limitation imposed by assuming a specific functional form, 
but the total number of time intervals (time resolution) is usually 
limited by the data. In practice, codes that focus on spectral 
fitting and stellar population parameters, 
such as STARLIGHT \citep{STARLIGHT} and pPXF \citep{Cappellari2004}, 
usually adopt the non-parametric approach, while the ones
that focus on constraining model parameters, such as 
CIGAL \citep{Noll2009,Boquien2019} and BEAGLE \citep{Chevallard2016},
prefer the parametric approach. Our analysis uses a stepwise model 
with 7 time intervals. As shown by both the mock tests 
and the fitting to real data, the stepwise model in 
general give results similar to the $\Gamma$+B model.
This consistency indicate that, as long as the SFH model is 
sufficiently flexible to describe the major stellar populations, 
the results from our spectral fitting are robust. 
For brevity, we will use the $\Gamma$+B model 
for the rest of our discussion.

\begin{figure}
\includegraphics[height=70mm]{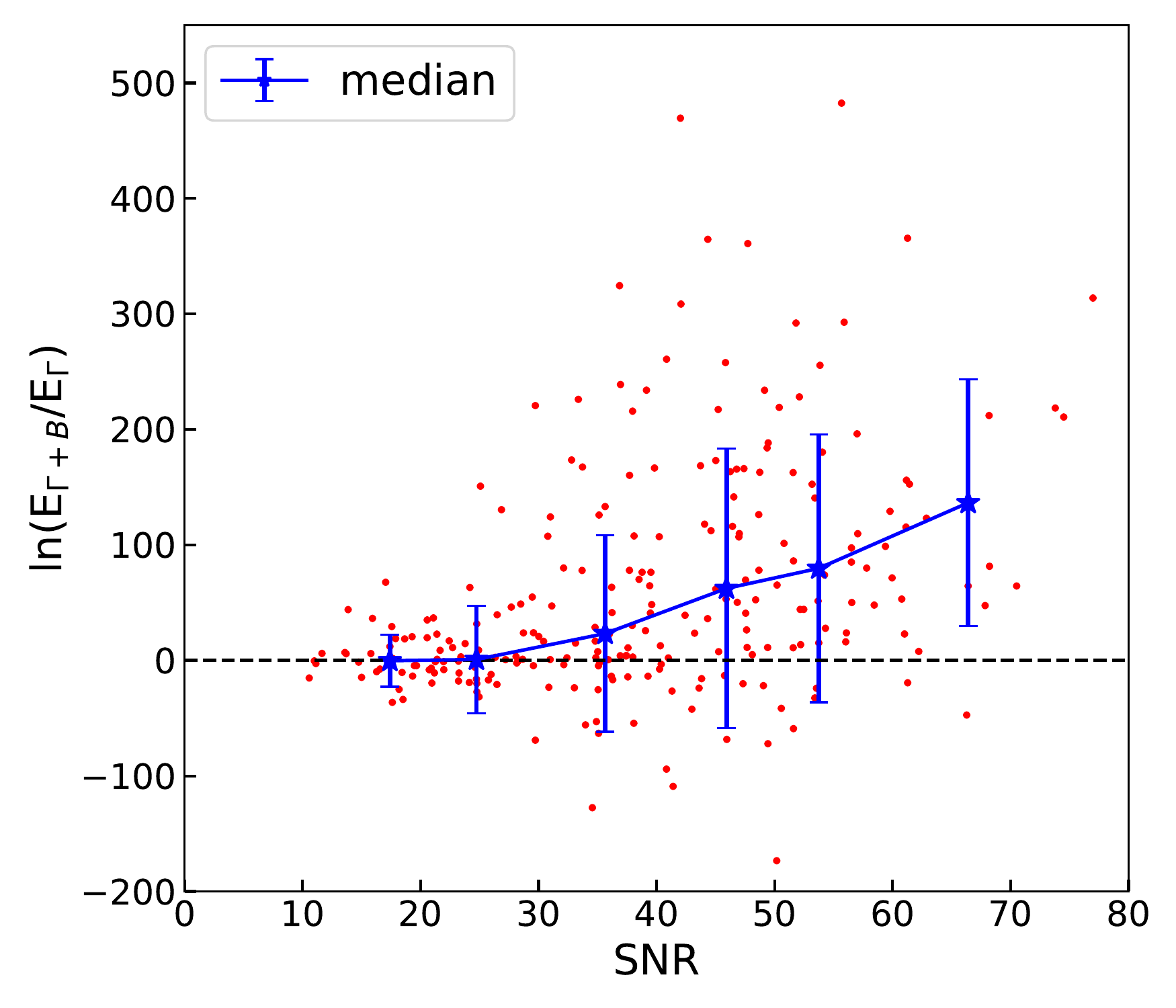}\\\
\includegraphics[height=70mm]{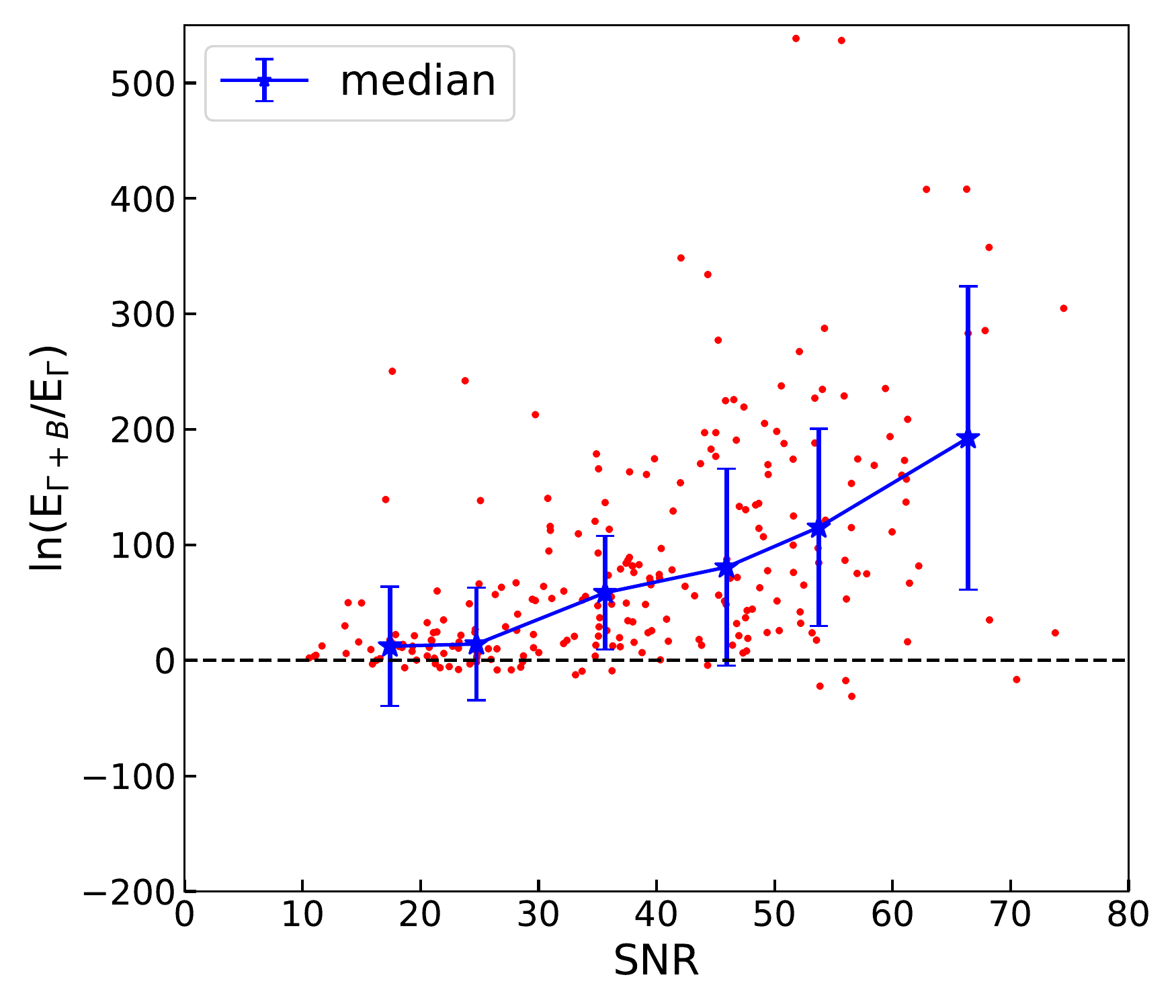}\\\
\caption{The evidence ratio between the $\Gamma$+B SFH and 
the $\Gamma$ SFH as a function of SNR of the stacked spectra, 
derived from the BC03 model (top panel) and the mixed model 
(bottom panel). Each red dot stands for the result of a MaNGA 
galaxy. Blue stars are the median values in five SNR bins 
connected by a blue line. Error bars are $1\sigma$ scatter 
of galaxies in individual bins.
}
\label{fig:evratio_bc}
\end{figure} 

\subsection{SSP models}
\label{ssec:discussion_ssp}

Spectral modelling is based on the linear combination of SSP templates. 
Thus, the accuracy and completeness of the SSP templates 
can influence fitting results. Unfortunately, the uncertainties 
of the SSP templates are not well-understood.  
As a test of this uncertainty, we use the BC03 model to perform a 
consistent check. In contrast to the E-MILES model, the BC03 SSPs 
are constructed with the STELIB \citep{Borgne2003} empirical 
stellar templates. Models assuming Padova isochrones \citep{Bertelli1994} 
and the Chabrier IMF are used in the comparison. 
The top panel of Fig.~\ref{fig:evratio_bc} shows the evidence ratio 
between the $\Gamma$+B and $\Gamma$ models, derived from the 
fitting with BC03 templates, as a function of the SNR. 
The trend seen here is similar to that shown in Fig.~\ref{fig:evratio}
although is slightly weaker.

We have also made a test using a mixture of E-MILES and BC03. 
In this model, the model fluxes are calculated as 
$F=F_{\rm BC}\times f_{BC}+F_{\rm EM}\times (1-f_{BC})$, 
where $F_{\rm BC}$ and $F_{\rm EM}$ are the fluxes obtained 
from the BC03 and E-MILES templates. The relative contribution of 
BC03 is described by $f_{BC}$, which is treated as a free parameter. 
The evidence ratio derived from this model, shown
in the bottom panel of Fig.~\ref{fig:evratio_bc},
is found to lie between the E-MILES and BC03 results. 

\begin{figure}
\centering
\includegraphics[height=0.45\textwidth]{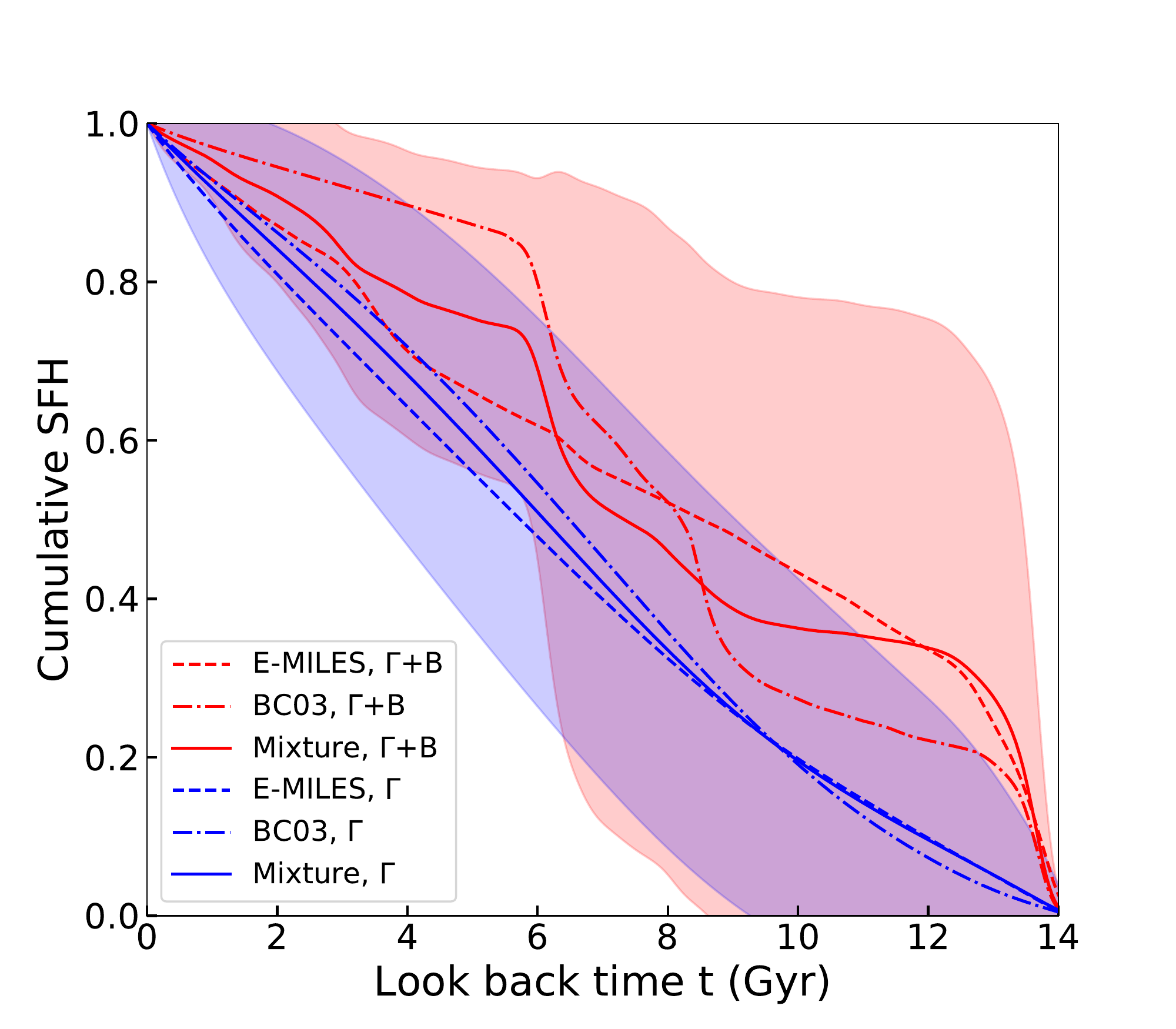}
\caption{
The cumulative SFH inferred from the best-fit SFH models using 
E-MILES templates (dash line), BC03 templates (dash-dotted line),
and the mixed model (solid line). Lines show the mean SFHs of sample, 
while the shaded region shows the scatter among sample galaxies 
obtained from the mixed model. 
Red and blue colours are for the $\Gamma$+B and $\Gamma$ models, respectively.
}
\label{fig:sfh_compare_bc}
\end{figure}

We plot the cumulative SFH inferred from the best-fit $\Gamma$+B model 
using E-MILES, BC03 and the mixture of the two in Fig.~\ref{fig:sfh_compare_bc}. 
The underestimation of the old stellar population by the 
simple $\Gamma$ SFH can be seen easily by comparing it with 
the $\Gamma$+B SFH, regardless of the assumed SSP model. 
For  $\Gamma$+B SFH, the inferred shapes of the SFH from the two 
SSP models are in overall 
agreement with each other. The prediction of the mixture model falls 
between the two; it is closer to E-MILES at early time  
and becomes closer to BC03 at later time.  
All the models reveal the presence of an old stellar 
population that dominates the stellar masses, 
but they differ in detail in their predictions
of the burst strength and age. 

These differences can be 
caused by the different constructions of the two SSP 
models, such as the underlying stellar templates, 
the isochrones used to calculate the SSPs, 
and even the methods used to populate stars in 
parameter space, but it is beyond the scope of the 
present paper to figure out the exact difference between 
the two SSP models. However, our test does show that 
the results obtained above are qualitatively robust 
against the variation of the SSP templates, but that 
quantitative details can be affected significantly 
by using different SSP models.

\subsection{The NIR photometry}
\label{ssec:discussion_nir}

\begin{figure}
\centering
\includegraphics[height=0.4\textwidth]{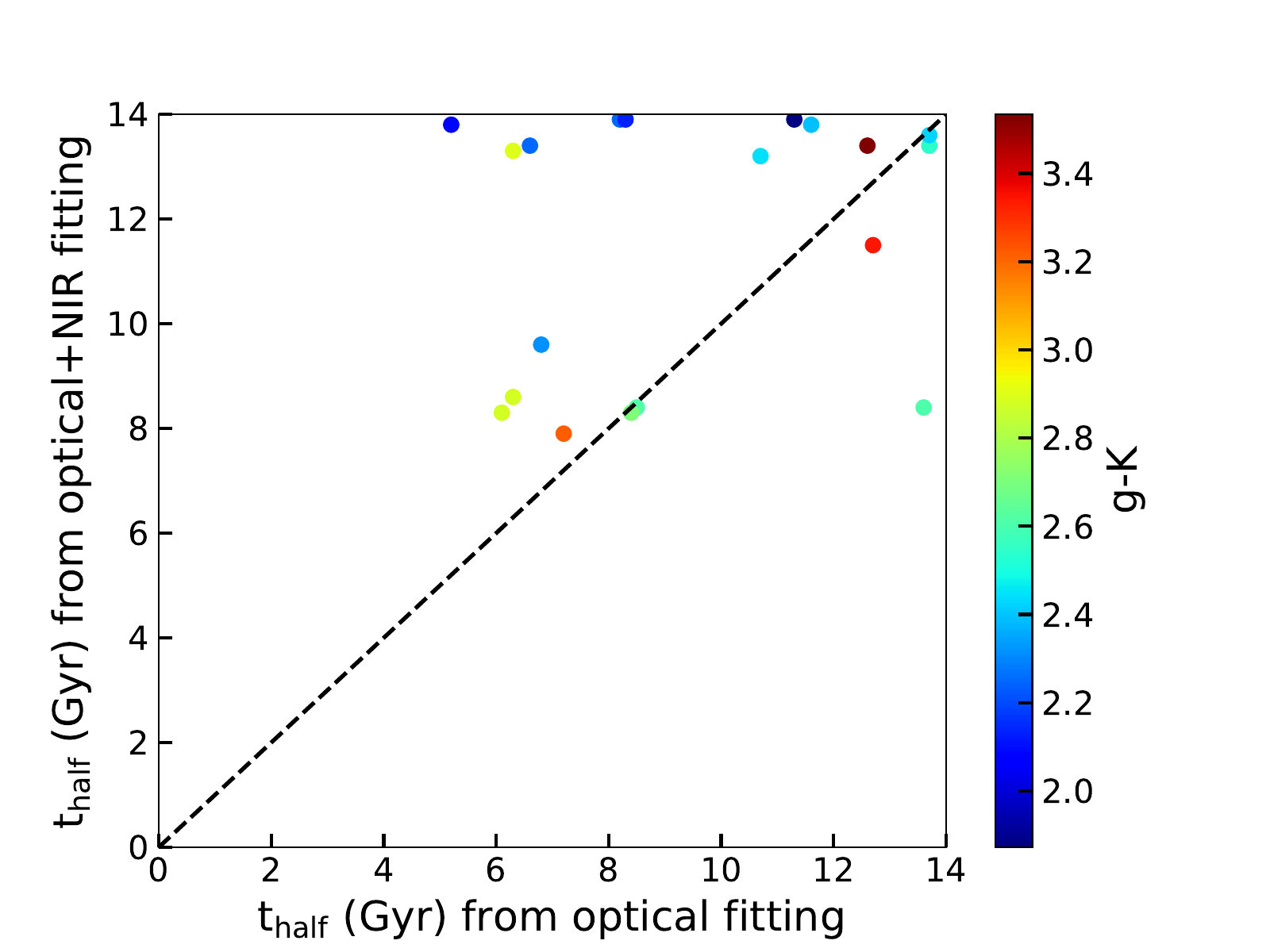}
\includegraphics[height=0.4\textwidth]{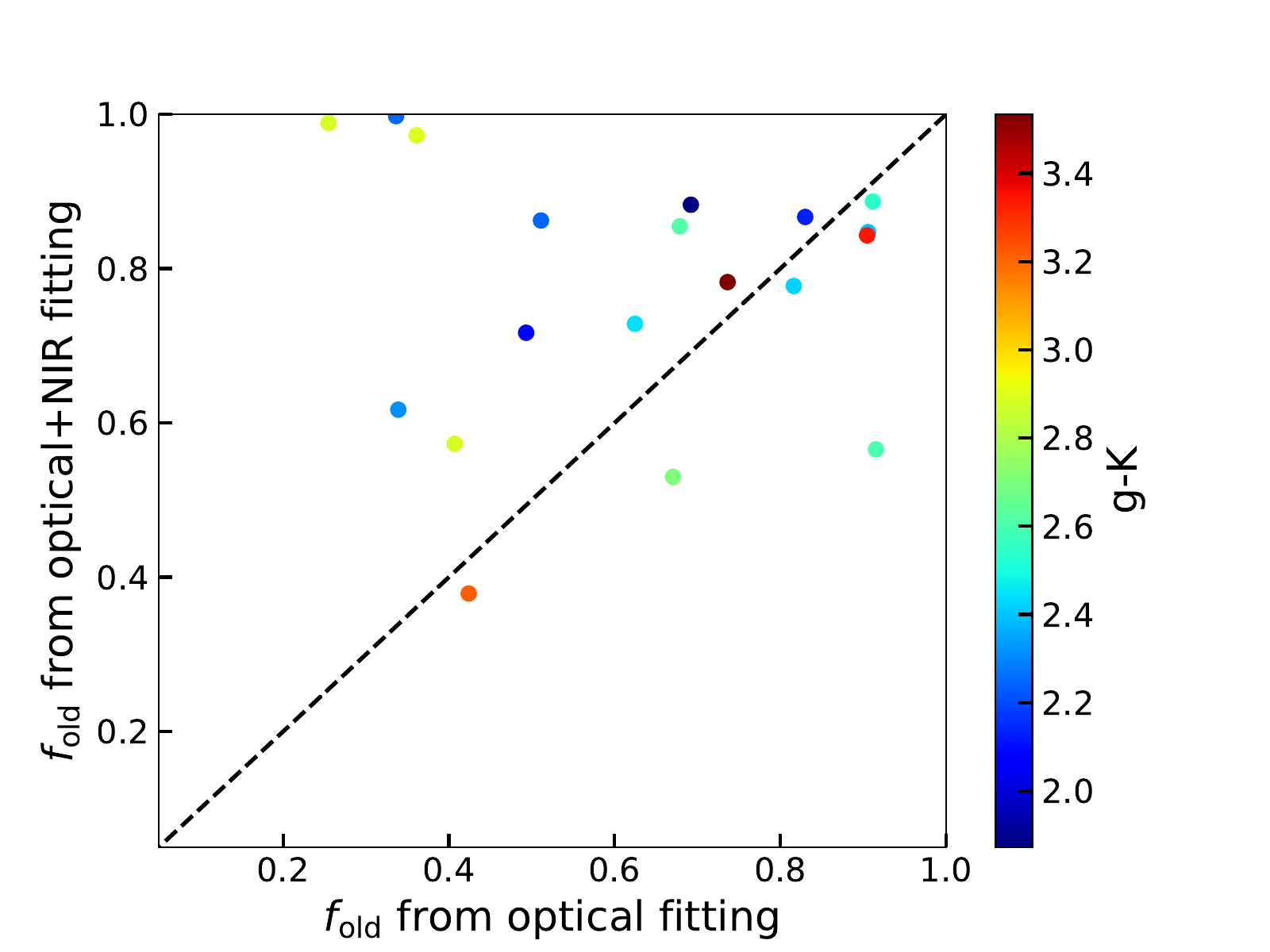}
\caption{Comparison of the half mass formation time (top) and the fraction 
of old population (bottom) obtained from fitting only the MaNGA stacked 
spectra (horizontal axis) and fitting both optical spectra and 
NIR photometry (vertical axis), using the BC03 templates. 
Results are colour coded by the $(g-K)$ colour of the galaxy.}
\label{fig:compare_nir_bc}
\end{figure}

The accuracy of the predicted NIR photometry depends on the 
accuracy of the SSP templates in the NIR. For the E-MILES model, 
the SSPs at $\lambda>8950$ {\rm \AA} are constructed 
using the empirical IRTF stellar template \citep{Cushing2005,Rayner2009}. 
This treatment provides a self-consistent E-MILES SSP spectra 
with moderately high resolution ($\sigma=60$\kms) at the NIR. 
In contrast, the BC03 model uses theoretical, low resolution 
NIR spectra of BaSel \citep{Westera2002}. This difference 
may affect the predicted NIR photometry. To test this uncertainty, 
we apply the BC03 model to the sample that has both MaNGA and 
UKIDSS data. We plot the half mass formation look-back time, 
$t_{\rm half}$, and the fraction of old (>8 Gyr) stellar 
population derived from optical only data and 
optical plus NIR data in Fig.~\ref{fig:compare_nir_bc}. 
Comparing with the results shown in Fig.~\ref{fig:compare_nir}, 
we see that the two SSP models reach the same conclusion 
that the inclusion of NIR data enhances the significance 
of the old stellar population.

In addition to the BC03 model, here we mention briefly another SSP model 
family, the SSP model of Maraston \citep{Maraston2005,Maraston2011}. 
This model was first presented in \cite{Maraston2005} using low-resolution 
theoretical BaSeL stellar libraries, and an updated version using a 
set of stellar templates was published in \cite{Maraston2011}. 
The Maraston model contains treatments of TP-AGB and HB stars
that are different from those in E-MILES and BC03.   
These treatments lead to redder colours for SSPs with ages around 1 Gyr. 
As the dwarf galaxies studied here have quite a significant population 
of such ages, the difference in the predicted 
$(g-K)$ colour can be as large as $\sim 0.15$.
Unfortunately, the Maraston model cannot be compared 
with E-MILES. The high-resolution model, which is needed for 
our modeling, is based on theoretical MARCS templates 
that have very limited age and metallicity coverages. 
The models based on empirical templates have relatively wider 
coverage in age and metallicity, but they generally only cover the 
optical range. Due to these reasons, it is difficult to make 
a meaningful comparison with this model.

In summary, our current SPS modelling is sufficient to describe the stellar 
population in the galaxies studied here, and our method is in general 
robust against the variation in both the SSP model and the SFH model. 
Thus, the main conclusions we have reached are robust, while 
the details are still uncertain.
 
\section{Summary and discussion}
\label{sec:summary}

In this paper, we analyse the spectra of a sample of low-mass galaxies using 
the Bayesian inference code BIGS. Our sample galaxies are selected from 
the SDSS IV-MaNGA IFU survey, and some of the galaxies also have 
NIR photometry obtained from the WSA catalogue. We stack the spatially 
resolved spectra of each galaxy within a radius of $1.0 R_e$  to 
obtain a representative high SNR spectrum for the galaxy. 
We also stacked spectra in three radial bins, $[0.0-0.3] R_e$, 
$[0.3-0.7] R_e$ and $[0.7-1.2] R_e$, to study possible radial 
variations. We analyse the stacked spectra using a full spectrum fitting 
approach, making use of the MaNGA spectra from 3400 \AA \ to 8900 \AA. 
In our analyses, we adopt the state-of-the-art E-MILES SSP templates, 
and assume different SFH models to derive the stellar 
population properties. We use Bayesian model selection to distinguish 
between different models, and use the posterior distribution to constrain 
model parameters. Our main results can be summarised as follows:
\begin{itemize}
\item
Based on Bayesian model selection, we demonstrate that low-mass 
galaxies contain an old stellar population that may be missed 
in results obtained from low-resolution spectra and a 
too restrictive model for the SFH.  
\item 
The best fit SFHs for both parametric and non-parametric 
models all show that most of the low-mass galaxies have formed more 
than half of their stellar mass at $z>2$. 
The half mass formation time and the cumulative SFH from 
our spectra fitting of the unresolved stellar populations are in 
good agreement with those obtained from resolved 
stars \citep{Weisz2011}. 
\item
The average mass fraction of the old stellar population derived from 
SFH models that are sufficiently flexible 
is as high as $\sim0.6$, while a simple $\Gamma$ model
significantly underestimates this fraction. 
This result is consistent with the resolved observation 
\citep{Weisz2011}, but inconsistent with some hydrodynamic 
simulations \citep[e..g.][]{Digby2019}.
\item
Central galaxies on average have more recent star formations
than satellite galaxies, indicating that star formation in 
satellite galaxies may be affected by their environment.
\item
Old stellar population in a low-mass galaxy 
exists not only in the central part, but is spread over the 
entire galaxy. On average, the variation of the 
the SFH with radius is rather weak.
\item
A model of SFH needs to be sufficiently flexible to reproduce 
an observed optical spectrum and the $(g-K)$ colour simultaneously. 
A higher fraction of the old stellar population is obtained 
when the $(g-K)$ colour is included as an additional constraint. 
However, the results obtained from the optical and NIR data 
have some tension, indicating that the current spectral 
synthesis model may not be sufficiently flexible to accomodate 
the data.
\item
We test potential uncertainties both in the SSP model and the SFH model, 
and find that our main results about the existence of an old stellar 
population in low-mass galaxies are robust. 
However, the details of the 
SFH are still poorly constrained.
\end{itemize}

Although SFHs in low-mass galaxies have been studied 
quite extensively, the underlying physical processes 
that regulate star formation are still poorly 
understood. For example, our results suggest that the SFH of 
low-mass galaxies may consist of an early star formation 
episode, where about half of the stellar mass was formed,  
followed by a secondary and more extended phase of star formation. 
This type of SFH is consistent with the empirical model 
of \cite{Lu2014}, which predicts that many dwarf galaxies 
have experienced a phase of rapid star formation 
at $z>2$. This enhanced star formation at $z>2$ may be 
related to the fast accretion of dark matter halos 
\citep[e.g.][]{Zhao2003,MoMao2004} and seems to be required 
by the observed upturn in the low-mass end of the 
stellar mass function of galaxies \cite[][]{Lan2016}, 
but is not predicted by many models of galaxy formation
\citep[][]{Lim2017}. In particular, the mass fractions in 
different stellar populations obtained from our analysis 
are not well reproduced by current hydrodynamic simulations
\citep[e.g.][]{Digby2019}. These suggest that the feedback 
effect assumed in the model to suppress star formation in 
low-mass halos may be too strong at high redshift. 
Clearly, the observational constraints on the SFH 
we obtain here can provide important information about 
the feedback processes operating in the population of low-mass 
galaxies.

Of all the approaches adopted to probe the SFH of low-mass galaxies,  
the most reliable methods are perhaps those based on resolved stars. 
However, such observations can be made only for a small number 
of nearby galaxies. In contrast, methods based on stellar 
population models can use a large sample of galaxies. 
Among the approaches based on spectral synthesis, 
SED fitting of broad-band photometry 
\citep[e.g.][]{Janowiecki2017,Telles2018} and absorption line analysis 
\citep[e.g.][]{Kauffmann2014} have been used to infer the SFHs of low-mass 
galaxies. Compared to these two approaches, our method based on 
full spectral fitting can in principle extract more information 
from the spectra. However, SED fitting has the advantage 
that muti-band photometry is easier to obtain. 
As we have shown, analysis based on spectra with 
limited wavelength coverage can result in biases in the inferred 
stellar population. Our Bayesian analysis that combines MaNGA 
optical spectra and NIR photometry from UKIDSS is an attempt to 
overcome this difficulty. As shown in \S\ref{ssec_nir}, this 
approach is promising in probing the stellar population in 
low-mass galaxies, in particular in revealing the old population 
that may be missed in earlier investigations.
However, it should also be noted that this approach is still in 
its early stage, and more explorations are needed to take full 
advantage of it. Accurate and self-consistent 
SSP templates are crucial for this type of analysis. 
As seen in \S\ref{sec:Uncertainties}, although variations 
in the SSP model generally do not change our results qualitatively, 
they do affect the details of inferences. 
Care must be taken in calibrating different observations. 
In the future, with improvements of our understanding about 
stellar spectra and of stellar spectral templates, 
the method and analysis proposed in this paper are 
expected to provide an important avenue to explore the star 
formation in low-mass galaxies.

\section*{Acknowledgements}
This work is supported by the National Key R\&D Program of China
 (grant No. 2018YFA0404502, 2018YFA0404503), and the National 
Science Foundation of China (grant Nos. 11821303, 11973030, 
11761131004, 11761141012). MB acknowledges FONDECYT regular grant 1170618. G.R. is supported by the National Research Foundation of Korea (NRF) through Grants No.
2017R1E1A1A01077508 and 2020R1A2C1005655 funded by the Korean Ministry of Education,
Science and Technology (MoEST), and by the faculty research fund of Sejong University. 

Funding for the Sloan Digital Sky Survey IV has been provided by the Alfred P.
Sloan Foundation, the U.S. Department of Energy Office of Science, and the Participating Institutions.
SDSS-IV acknowledges support and resources from the Center for High-Performance Computing at
the University of Utah. The SDSS web site is www.sdss.org.
\par
SDSS-IV is managed by the Astrophysical Research Consortium for the Participating Institutions of the SDSS Collaboration including the Brazilian Participation Group, the Carnegie Institution for Science, Carnegie Mellon University, the Chilean Participation Group, the French Participation Group, Harvard-Smithsonian Center for Astrophysics, Instituto de Astrof\'isica de Canarias, The Johns Hopkins University, Kavli Institute for the Physics and Mathematics of the Universe (IPMU) / University of Tokyo, Lawrence Berkeley National Laboratory, Leibniz Institut f\"ur Astrophysik Potsdam (AIP), Max-Planck-Institut f\"ur Astronomie (MPIA Heidelberg), Max-Planck-Institut f\"ur Astrophysik (MPA Garching), Max-Planck-Institut f\"ur Extraterrestrische Physik (MPE), National Astronomical Observatories of China, New Mexico State University, New York University, University of Notre Dame, Observat\'ario Nacional / MCTI, The Ohio State University, Pennsylvania State University, Shanghai Astronomical Observatory, United Kingdom Participation Group, Universidad Nacional Aut\'onoma de M\'exico, University of Arizona, University of colourado Boulder, University of Oxford, University of Portsmouth, University of Utah, University of Virginia, University of Washington, University of Wisconsin, Vanderbilt University, and Yale University.

\section*{Data availability}
The data underlying this article were accessed from: SDSS DR15 \url{https://www.sdss.org/dr15/}; UKIDSS  \url{http://wsa.roe.ac.uk/}. The derived data generated in this research will be shared on reasonable request to the corresponding author.
\bibliographystyle{mnras}
\bibliography{SFH} 

\bsp	
\label{lastpage}

\appendix
\section{Posterior predictive distribution of the $(g-K)$ colour}
\label{appedix_nir}

In Bayesian context, once the posterior distribution $P(\theta|D,H)$ 
is obtained through a chosen posterior sampler, one can make predictions 
by marginalizing the desired likelihood function over the posterior 
distribution. For example, the 
posterior predictive distribution (PPD) of a set of observable quantities 
$D'$, given the constraining data $D$, can be written as 
\begin{equation}
P(D'|D,H)=\int P(D'|\theta, H) P(\theta|D, H)\,d\theta
\label{eq:PPD}
\end{equation}
where $P(D'|\theta, H)$ is the desired likelihood function describing 
the probability distribution of $D'$ expected from the model $H$ specified by 
the parameter set $\theta$. Note that the set of the predicted observable, $D'$, 
is not necessarily related to the constraining data $D$: $D'$ can be anything 
that can be predicted by the model through $P(D'|\theta, H)$. 
For a complex model $H$, such as the one we are concerned here, the PPD cannot 
be obtained analytically. In this case, one can select a large sample of models, 
$\{\theta_l\} (l=1, 2\cdot\cdot\cdot L)$ (with $L$ the sample size), 
from the posterior distribution, make a prediction for $D'$ using each of the $\theta 
\in \{\theta_l\}$, and obtain a sample of the PPD, $P(D'|D, H)$, from the 
sample of $D'$ obtained from $\{\theta_l\}$. The PPD so obtained can then be compared 
with the observational data of $D'$ to check whether the original model 
can accommodate the observational data. 

Once the PPD is obtained, one can check a specific model family using a 
procedure called the posterior predictive check (PPC). The basic idea is 
the following: if the model is correct then the data replicated from the model
should have a distribution that can accommodate the observational data 
of $D'$, which will be denoted as ${\cal D}'$. Any significant discrepancy 
between the PPD and ${\cal D}'$ will signify the inadequacy of the 
hypothesis $H$. To compare the PPD with the data, one can define a test 
statistic $T(D')$. The tail probability (the $p$-value) of 
the test statistic can then be used to assess the `goodness of fit' 
of the model to the data. In the Bayesian context, the p-value can be 
defined as
\begin{eqnarray}\label{eq_pvalue}
p&=&P[T(D')\geq T({\cal D}'| D)]\nonumber\\
   &=&\int \int I_{T(D')\geq T({\cal D'})}
   P(D'|\theta, H)P(\theta|D,H)\,d\theta d D'\,,
\end{eqnarray}
where $I_q$ is the indication function for the condition $q$
($I_q=1$ if $q$ is true and 0 otherwise). If the observational data 
${\cal D}'$ is incompatible with the model, then the test statistic 
from the data,  $T ({\cal D}')$, should be a significant outlier of 
the distribution of $T (D')$ predicted by the model. In practice, if 
the posterior predictive p-value is close to 0 or 1, then the model 
is most likely inadequate.

\begin{figure*}
\centering
\includegraphics[height=0.4\textwidth]{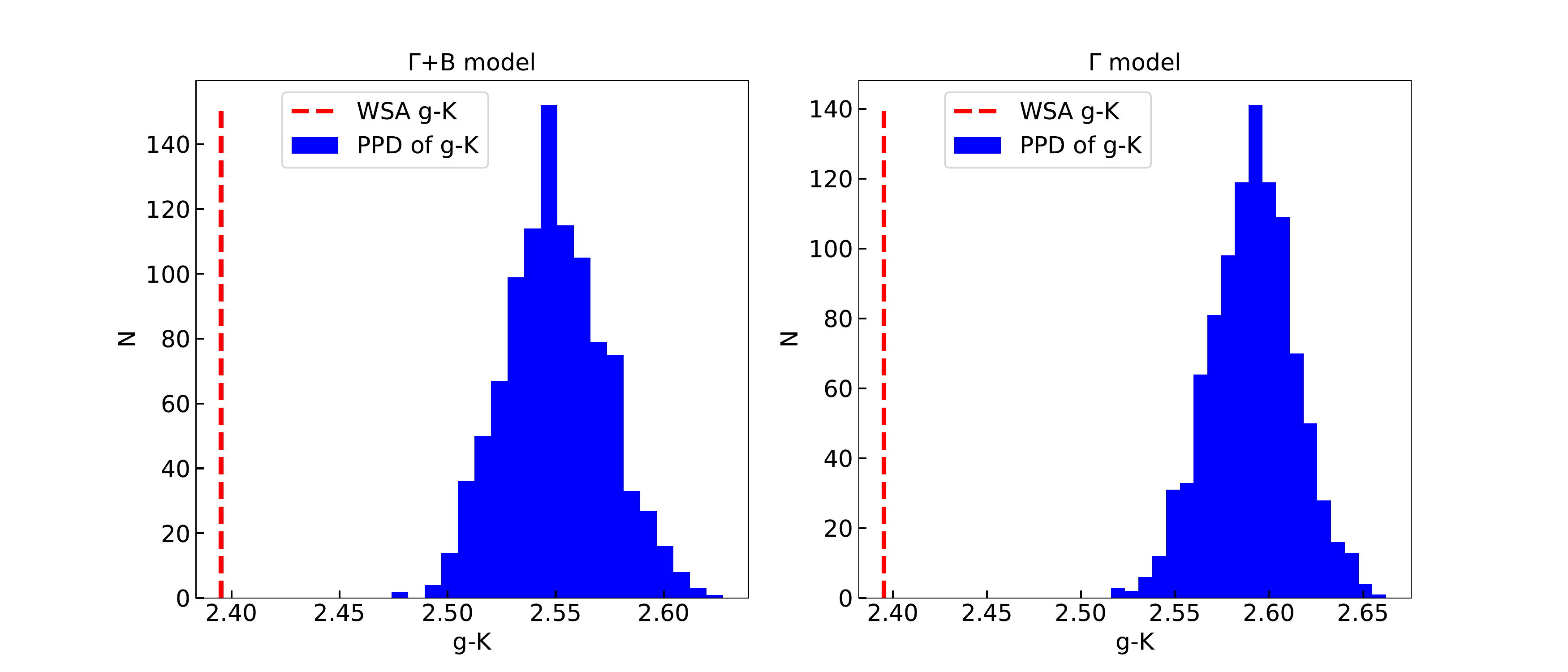}
\includegraphics[height=0.4\textwidth]{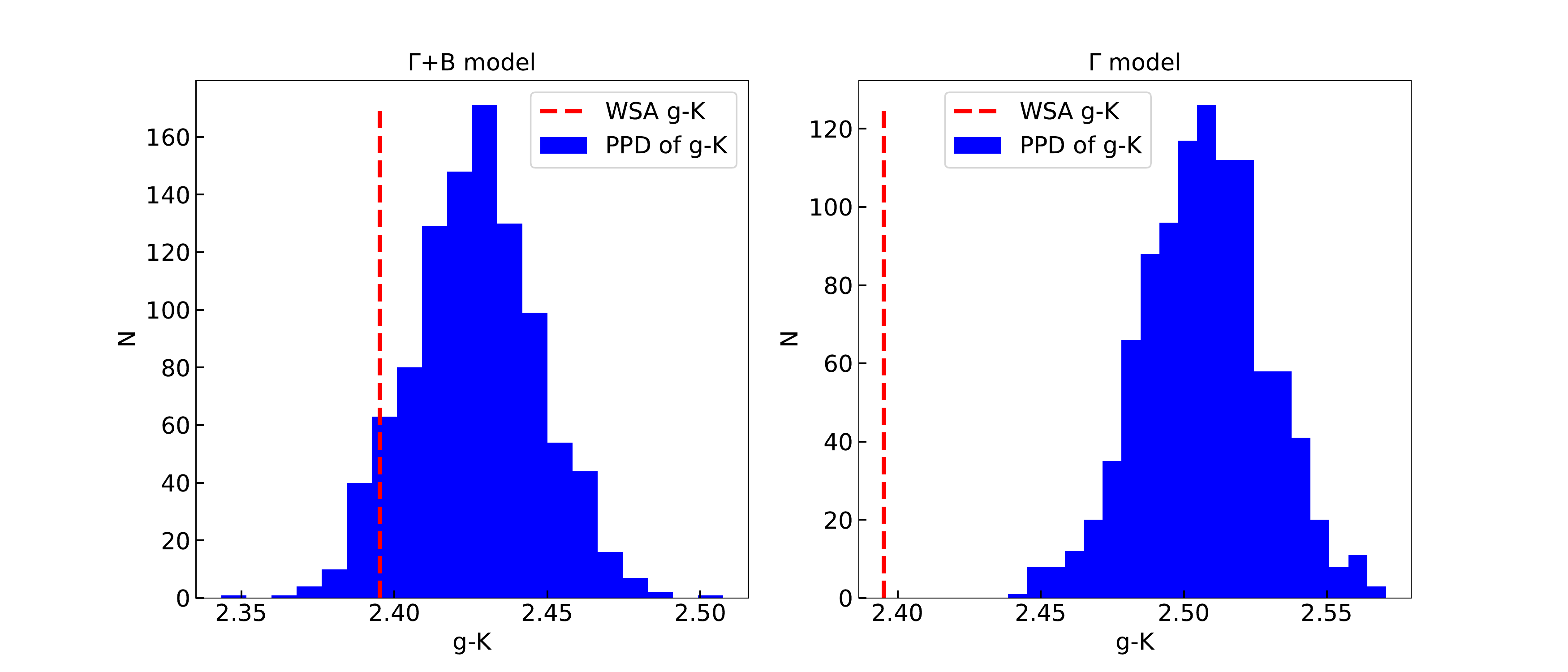}
\caption{Histograms showing the posterior predictive distribution of the 
$(g-K)$ colour. Left panels are results obtained using the 
$\Gamma$+B model, while right panels are using the $\Gamma$ model. 
Top panels are based on MaNGA optical spectra, while bottom panels 
are results that include the $(g-K)$ colour. Red dash line in each panel 
denotes the $(g-K)$ colour from UKIDSS.}
\label{fig:PPD}
\end{figure*}

Here We use the models constrained by the MaNGA data to 
make predictions for the NIR photometry and check whether or not 
the models can also accommodate the NIR data. We use the fitting of 
galaxy 9490-6104 as an example.  
To get the PPD, we draw $L=1000$ samples from the posterior 
distribution obtained from the MaNGA spectra of this galaxy    
using the $\Gamma$ model and $\Gamma$+B model, respectively. 
We then use the $(g-K)$ colour as the predicted observable $D'$.
The PPDs derived from the two SFH models are shown in 
Fig.~\ref{fig:PPD}. The PPD of the $\Gamma$ model peaks at 
$(g-K)=2.61$ while that of the $\Gamma$+B model 
at $(g-K)=2.56$. The measurement from WSA for this galaxy is
$(g-K)=2.39$. These results indicate that even with the inclusion 
of an early burst population, the lack of constraints from NIR bands 
may still lead to some bias in the stellar population.

\begin{figure*}
\centering
\includegraphics[height=0.4\textwidth]{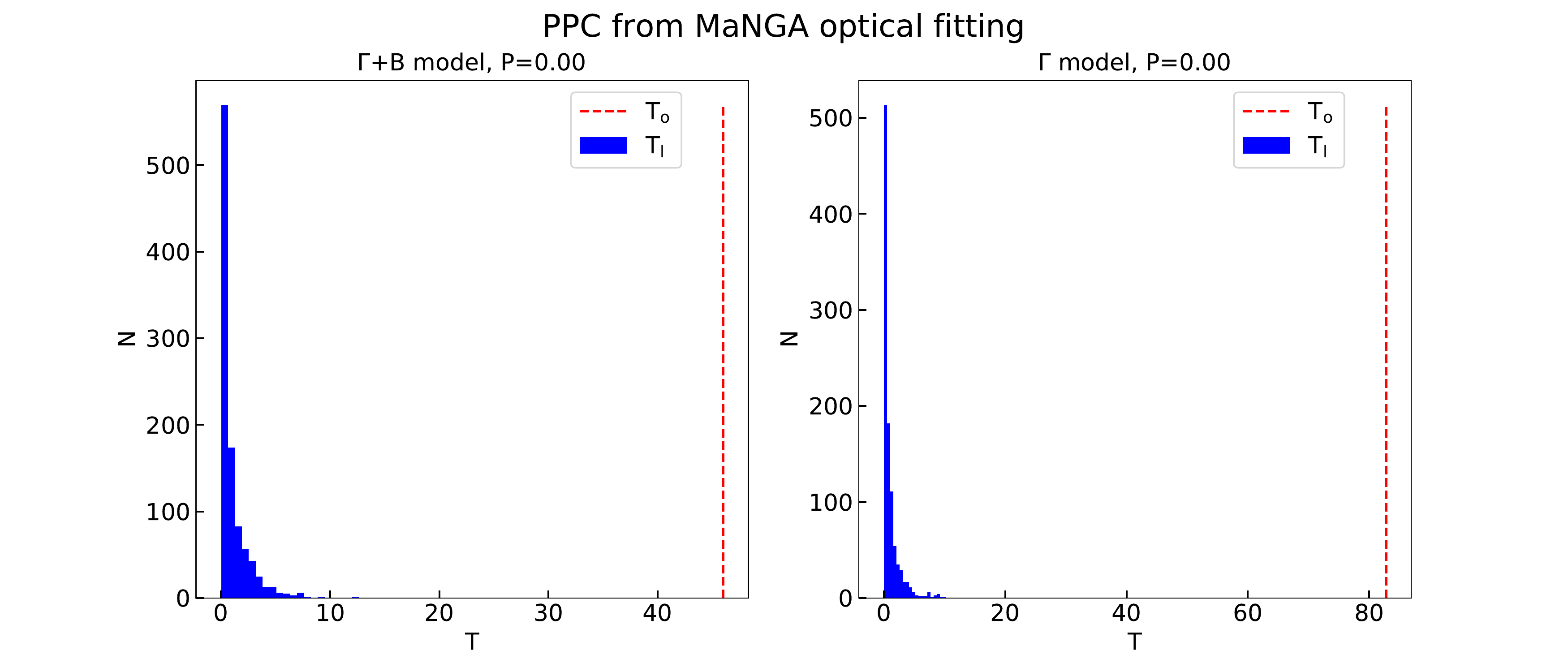}
\includegraphics[height=0.4\textwidth]{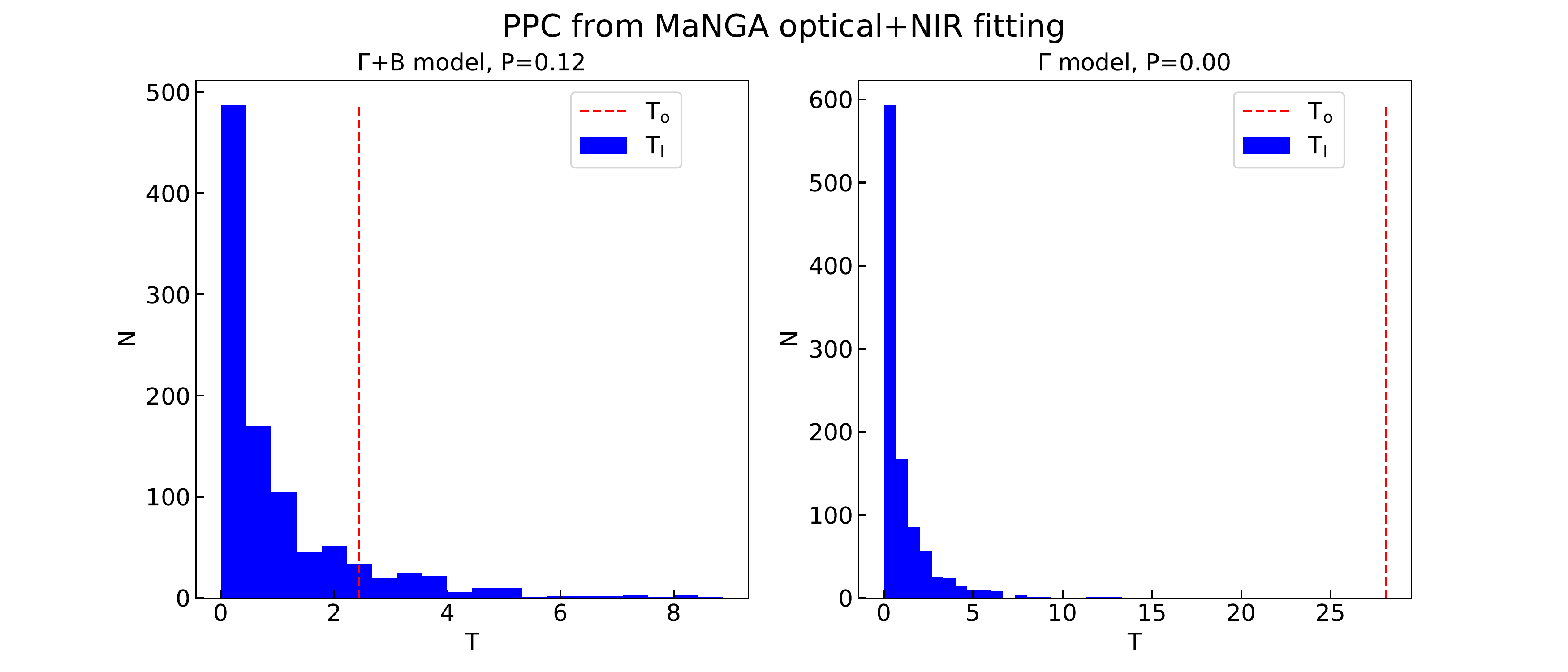}
\caption{Posterior predictive check for the $\Gamma$+B model (left) 
and the $\Gamma$ model (right). 
Histograms show the distribution 
of the test statistics T${\rm _l}$ for the 1000 samples drawn from the 
posterior distribution. Top panels are results based on fitting the MaNGA optical 
spectra, while bottom panels are results that include the $(g-K)$ colour. 
Red dash line in each panel denotes the test statistics T${\rm _o}$ calculated 
from the UKIDSS observation.}
\label{fig:PPC}
\end{figure*}

As we have chosen the $(g-K)$ colour (denoted as $K'$ for convenience) 
as our test observable, we can use a $\chi ^2 $-like test statistic to 
perform the PPC on the fitting results. The corresponding 
test statistic can be written as 
\begin{equation}
T_l(K_l')={(K_l'-\overline{K}')^2 \over \sigma'^2}
\label{T_chi2}
\end{equation}
where  $\overline{K}'$ and $\sigma'$ are, respectively, the mean and 
standard deviation of the PPD obtained from a posterior sample. 
For an observed $(g-K)$ colour, $K_o'$, the corresponding statistic is 
\begin{equation}
T_o={(K_o'-\overline{K}')^2
\over \sigma'^2}\,.
\label{TO_chi2}
\end{equation}
The $p$-value defined by equation (\ref{eq_pvalue}) can then be obtained. 
Fig.~\ref{fig:PPC} shows the distribution of the test statistics 
$T$ and the $T_o$ obtained from the observed $(g-K)$ colour, 
together with the $p$-value obtained from equation (\ref{eq_pvalue}). 
The $p$-value is $0.0$ in both of the two fitting results, indicating 
that the models are over-constrained by the MaNGA spectra.   

As a comparison, we can make inferences by including the observed 
$(g-K)$ colour as a constraint. Using the likelihood function in 
equation (\ref{eq:likelyhood_NIR}), we apply the posterior predictive 
check method to the fitting results. The results are shown 
in the bottom panels of Fig.~\ref{fig:PPD} and Fig.~\ref{fig:PPC}, 
respectively. By including this additional constraint, the 
$\Gamma$+B model can now pass the PPC test, while the $\Gamma$ model
still fail to reproduce the observation.

\begin{figure*}
\centering
\includegraphics[height=0.35\textwidth]{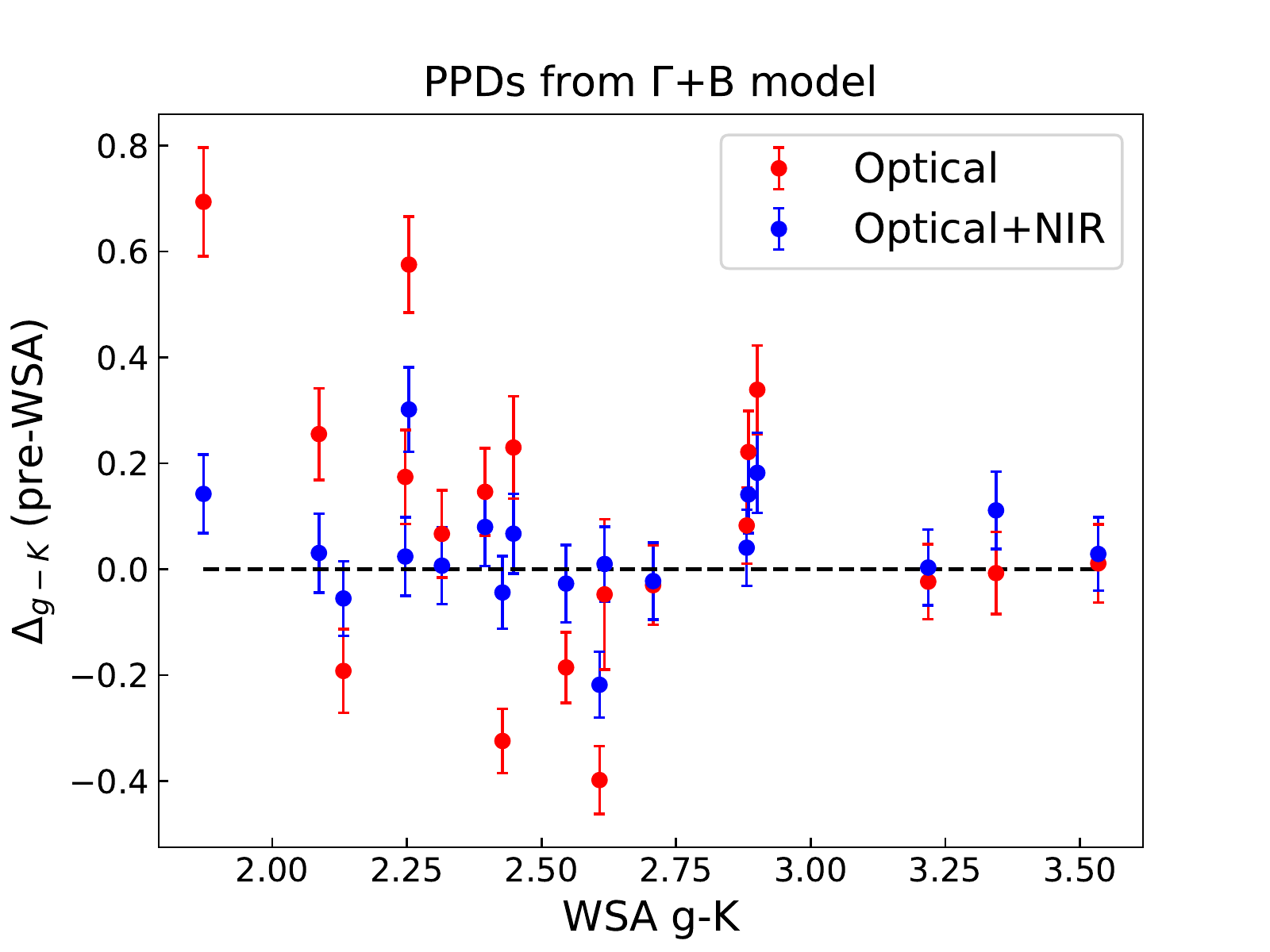}
\includegraphics[height=0.35\textwidth]{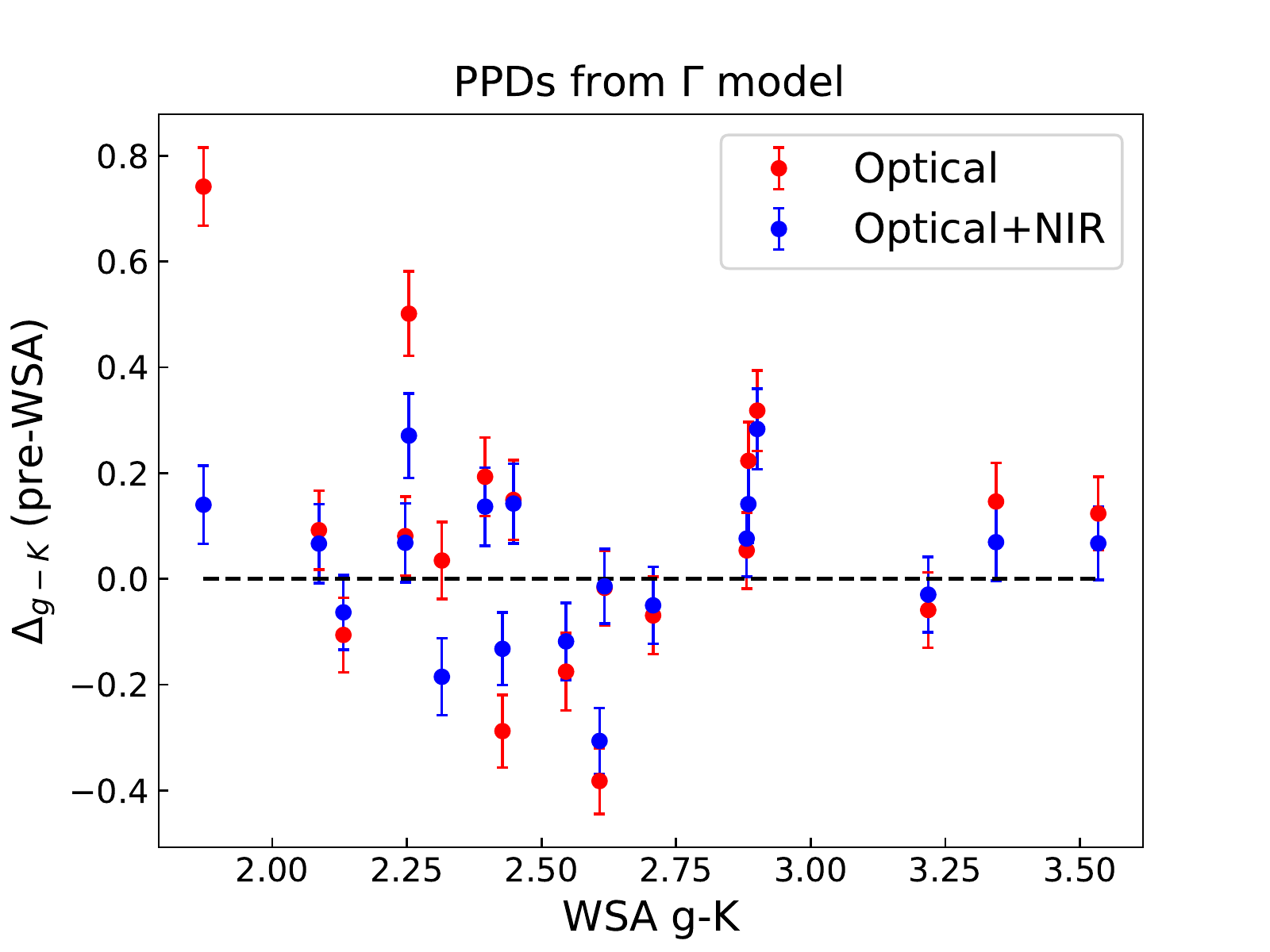}
\caption{Differences between the posterior predicted $(g-K)$ colour 
from fitting MaNGA optical spectra and the WSA values. Results are 
derived from $\Gamma$+B model (left) and $\Gamma$ model (right), red and 
blue dots are from MaNGA optical fitting and optical+NIR fitting, 
respectively. Error bars denote the 3$\sigma$ level of the PPD.}
\label{fig:PPC_sample}
\end{figure*}

To statistically characterize the results, we apply PPC to all the 
19 galaxies that have UKIDSS NIR measurements. 
We derive the PPD from the posterior distributions of the fitting results 
and calculate the $3\sigma$ level of the distribution as an approximate 
boundary. Successful models should predict a PPD that covers the probable 
observational ranges, and so the observation data are expected 
to fall within the boundary. We plot the difference between the predicted 
and observational $(g-K)$ colours, denoted as  $\Delta_{\rm g-K}$, in 
Fig.~\ref{fig:PPC_sample}, and show the corresponding $3\sigma$ boundary 
as error bars. As one can see, if only the MaGNA spectra are used in the 
fitting, both the SFH models fail to pass the PPC test in many case. 
In contrast, if the NIR data is included in the fitting, 
the $\Gamma$+B model can generally pass this test, while the $\Gamma$ 
still fails to reproduce the observation in many cases
(see blue points in Fig.~\ref{fig:PPC_sample}). 
These results strengthen our conclusion that the $\Gamma$+B model 
is a better description of the true SFH. They also indicate the importance 
of using the NIR data in order to correctly model the stellar population. 

However, those results also signify tensions between the predictions based 
on high-resolution spectra with limited wavelength coverage and those 
based on broadband photometry covering a larger wavelength range.  
Ideally, if the optical spectra do not provide tight 
constraints on the stellar populations that can affect the NIR 
fluxes significantly, the posterior distribution inferred  
from the optical data should be broad enough to accommodate the 
observed NIR fluxes. In reality, however, the situation may be 
more complicated. Spectral synthesis modeling of observed spectra 
is high dimensional, and the posterior distribution may be complex. 
If the model cannot describe the observed spectra perfectly, 
there is no guarantee that the 'best-fit' model derived from 
part of a spectrum is also the best-fit model for the entire spectrum. 
The posterior distribution inferred from part of the spectrum 
may then lead to biased predictions for the spectrum outside the 
observed window. In principle, any uncertainties in the model itself 
should be included in the likelihood function. Unfortunately, such 
uncertainties are difficult to quantify for the spectral synthesis 
model concerned here.

\begin{figure}
\centering
\includegraphics[height=0.35
\textwidth]{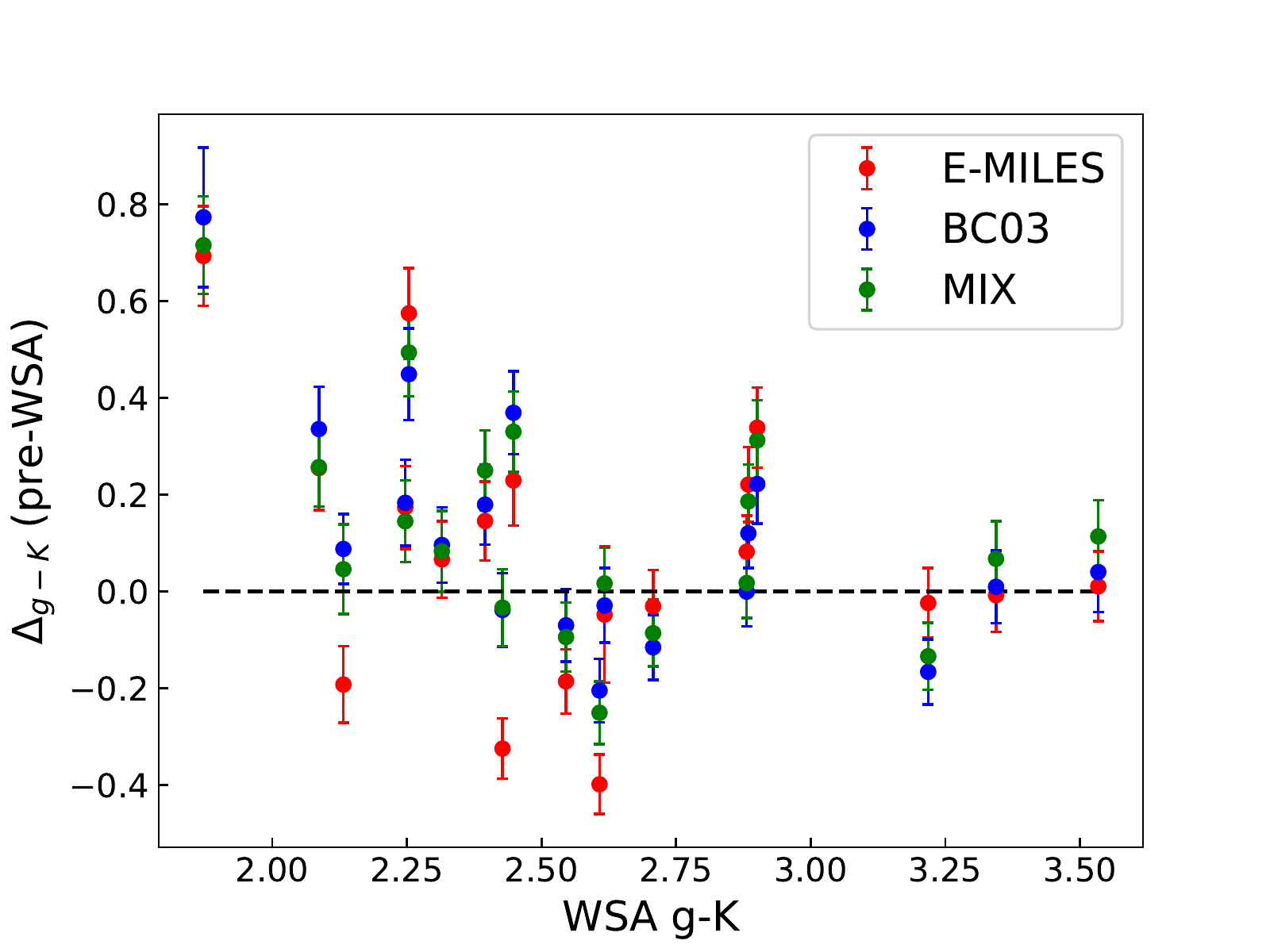}
\caption{Differences between the posterior predicted $(g-K)$ 
colour from fitting MaNGA optical spectra and the WSA values. 
Results are derived from E-MILES (red), BC03 (blue), and mixed models (green). 
Error bars are the $3\sigma$ level of the PPD.}
\label{fig:PPC_templates}
\end{figure}

 Uncertainties of the SSP templates are perhaps the most 
important in affecting the spectral synthesis model.   
Available Models based on different SSP templates, with their 
own merits and shortcomings, may work better in some 
cases but worse in others. To test this, we use our fitting results 
based on the BC03 model and the mixture model described in 
\S\ref{ssec:discussion_ssp}. The $\Delta_{\rm g-K}$ predicted 
by these two models are shown in Fig.~\ref{fig:PPC_templates}
in comparison with those predicted by the E-MILES model. 
As we can see, the BC03 model can reach consistencies between 
the optical fitting and the NIR observations in some cases
where E-MILES fails, and vise versa. 

In summary,  the conflict between optical predictions and 
the NIR constraints may indicate that the current model is not flexible 
enough. Allowing flexibility in, e.g. SSP templates, may alleviate 
such tension. As the factors that can contribute to the conflict 
are difficult to quantify, making full use of all available data 
to constrain model parameters may be a reasonable approach to reach 
a compromise among different constraints.

\end{document}